\documentclass[pra,superscriptaddress,twocolumn]{revtex4}
\usepackage{graphicx}
\usepackage{amsfonts,amsmath,amssymb,float}
\usepackage{bm,dsfont}
\usepackage{color}
\usepackage{bbm}
\usepackage{longtable}
\usepackage[dvips]{epsfig}
\usepackage{amsmath,amssymb,lscape,float}
\usepackage{hyperref}
\usepackage{listings}

\newcommand{\ket}[1]{|#1 \rangle}

\newcommand{\uber}[2]{{{#1}\choose{#2}}}

\begin{document}
\title{Harnessing subspace controllability: Dynamical  generation of Dicke states in 
Heisenberg-coupled qubit arrays with a single local control}

\author{Vladimir M. Stojanovi\'c}
\affiliation{Institut f\"{u}r Angewandte Physik, Technical University of Darmstadt, D-64289 Darmstadt, Germany}

\author{Tommaso Calarco}
\affiliation{Forschungszentrum Jülich GmbH, Peter Grünberg Institute, Quantum Control (PGI-8), 52425 Jülich, Germany}
\affiliation{Institute for Theoretical Physics, University of Cologne, Zülpicher Straße 77, 50937 Cologne, Germany}
\affiliation{Dipartimento di Fisica e Astronomia, Università di Bologna, 40127 Bologna, Italy}

\author{Andrea Muratori}
\affiliation{Dipartimento di Fisica e Astronomia, Università di Bologna, 40127 Bologna, Italy}

\date{\today}

\begin{abstract} 
We explore the feasibility of realizing Dicke states in qubit arrays with always-on isotropic Heisenberg coupling between adjacent qubits, assuming a single Zeeman-type control acting in the $z$ direction on an actuator qubit. The Lie-algebraic criteria of controllability imply that such an array is not completely controllable, but satisfies the conditions for subspace controllability on any subspace with a fixed number of excitations. Therefore, a qubit array described by the model under consideration is state-to-state controllable for an arbitrary choice of initial and final states that have the same Hamming weight. This limited controllability is exploited here for the time-efficient dynamical generation of an $a$-excitation Dicke state $|D^{N}_{a}\rangle$ ($a=1,2,\ldots, N-1$) in a linear
array with $N$ qubits starting from a generic Hamming-weight-$a$ product state. To dynamically generate the desired Dicke states -- including 
$W$ states  $|W_{N}\rangle$ 
as their special ($a=1$) case -- in the shortest possible time with a single local $Z$ control, we employ an optimal-control scheme based on the {\em dressed chopped random basis} (dCRAB) algorithm. We optimize the target-state fidelity over the expansion coefficients of smoothly-varying control fields in a truncated random Fourier basis; this is done by combining Nelder-Mead-type local optimizations with the multistart-based clustering algorithm that facilitates searches for global extrema. In this manner, we obtain the optimal control fields for Dicke-state preparation in arrays with up to $9$ qubits. Based on our numerical results, we find that the shortest possible state-preparation times scale as $\mathcal{O}(N^{2.08})$ for $W$ states and $\mathcal{O}(N^{1.78})$ for $a=2$ Dicke states. Finally, we demonstrate the robustness of our 
dCRAB-based state-engineering scheme against various types of imperfections of relevance for its anticipated experimental implementation.
\end{abstract}

\maketitle
\section{Introduction}
Introduced in the context of the phenomenon of 
superradiance~\cite{Dicke:54}, Dicke states have in recent years been extensively investigated in connection with a variety of emerging quantum-technology applications. To be more specific, favorable properties of this class of highly-entangled multiqubit states~\cite{Bergmann+Guehne:13} -- above all, their extreme robustness to particle loss~\cite{Neven+:18,Zhang+:25}, 
their immunity to collective dephasing noise~\cite{Lidar+Whaley:03},
as well as their permutationally-symmetric character~\cite{Stojanovic+Nauth:23} -- make them a viable candidate for applications 
in areas as diverse as quantum game theory~\cite{Ozdemir+:07}, quantum networking~\cite{Prevedel+:09}, quantum metrology~\cite{Toth:12,Saleem+:24}, quantum error correction~\cite{Ouyang:14}, and quantum 
combinatorial optimization~\cite{Golden+:21}, to name but a few.

Motivated by their  anticipated promise in the realm 
of quantum technology, Dicke states 
have already been realized in some classes of physical systems, e.g. photonic ones~\cite{Prevedel+:09,Wieczorek+:09}. In addition,
a multitude of proposals for the realization of those states in diverse physical systems were put forward in the past; examples are 
furnished by trapped ions~\cite{Hume+:09,Lamata+:13}, neutral atoms~\cite{Shao+:10,Srivastava+:26}, 
spin-$1/2$ ensembles~\cite{Zhang+Jing:26}, and superconducting qubits~\cite{Wu+:17}.
It should be stressed, however, that most of the heretofore demonstrated schemes for the realization of Dicke states -- as well as the existing theoretical proposals thereof -- allow one to engineer only particular instances of those states, rather than an arbitrary state
from this family~\cite{Kobayashi+:14}. Furthermore, most of the envisioned state-preparation schemes are fraught with additional limitations  -- for example, they require specific initial states, individual qubit addressing, or fixed (pre-determined) qubit-array topologies~\cite{Albarran+:26} as a prerequisite for realizing Dicke states.

In this paper, the feasibility of the dynamical generation of Dicke states 
starting from an initial product state is investigated in the special case of linear qubit arrays with nearest-neighbor Heisenberg-type coupling and a single local control. This nontrivial quantum-state-engineering 
problem is addressed here by making use 
of the Lie-algebraic controllability criteria~\cite{Jurdjevic+Sussmann:72,
Huang+Tarn+Clark:83,Ramakrishna+Rabitz:96,
D'AlessandroBook} and 
employing the optimal-control methods~\cite{Mueller+:22,Ansel+:24,ControlReview:25}. While the former guarantee the existence of time-dependent control fields that render the envisioned state-engineering scheme possible -- but without specifying the actual time dependence of those fields -- the latter can be utilized to determine the sought-after time dependence~\cite{Cappellaro+:07,
Stefanescu+:25,deLima+:26}.

It should be stressed that -- unlike in most of the previously proposed schemes for the preparation of Dicke states -- the present work corresponds to the scenario with an {\em always-on} coupling between qubits~\cite{Benjamin+Bose:03}. Therefore, before even embarking in earnest on a discussion of controllability in qubit arrays with always-on Heisenberg-type coupling~\cite{Heule+:10} it is pertinent to first elaborate -- from a more general perspective -- on the role 
and utility of always-on couplings in the field of quantum computing (QC), with emphasis on solid-state QC platforms. 

The tantalizing recent progress achieved in various physical platforms for QC has relied, above all, on the 
ever-increasing accuracy of realizing entangling two-qubit gates~\cite{Chen+:25}. These high-fidelity experimental realizations 
have been based on tunable qubit-qubit
couplings, which are switched
on only when a two-qubit gate is required. In particular, in solid-state QC platforms tunable couplers -- which engender such 
switchable couplings -- usually entail multiple circuit elements, thus necessitating their 
own external control for tuning the
coupling~\cite{Chen+:14}; this leads to overheads in fabrication and wiring, 
adding further technological complexity in the process of scaling up the device. By
contrast to tunable couplers, their counterparts that give rise to fixed (always-on) couplings do not require such additional components, thus allowing for a considerable simplification of the hardware architecture and alleviating the inherent difficulties of scaling up.

It is well-understood by now that the Heisenberg interaction between qubits by 
itself does not suffice for universal quantum computation (UQC)~\cite{DiVincenzo+:00},
by contrast to the Ising- and $XY$-type interactions~\cite{NielsenChuangBook}. It 
has been demonstrated, however, that UQC can still be realized with the Heisenberg interaction alone if encoded qubit states -- with the role of logical qubits being played by triples~\cite{DiVincenzo+:00} or pairs~\cite{Levy:02} of physical qubits -- are introduced, leading to the concept of encoded universality~\cite{EncodedUniversality}. Another useful insight as to the potential relevance of this type of qubit-qubit coupling -- which was also discussed in the context of measurement-based QC~\cite{Tanamoto+:12} and stabilizer codes~\cite{Tanamoto+:13} -- within the circuit model of QC came from Lie-algebraic studies of interacting spin-$1/2$ chains acted upon by time-dependent control fields~\cite{Schirmer++:08,
Wang++:16}. To be more specific, it was shown that a spin-$1/2$ chain with an always-on Heisenberg-type interaction~\cite{Benjamin+Bose:03} is completely 
controllable provided that at least two noncommuting controls -- e.g., a Zeeman-type magnetic field 
with nonzero components in two mutually 
orthogonal directions (e.g., $x$ and $y$) 
that act on a single spin (qubit)~\cite{Wang++:16}. 
In other words, under this last condition, an arbitrary (multiqubit) quantum gate can 
in principle be realized in such 
an array~\cite{Heule+:10,
Heule+:11,Stojanovic:19}, which 
is equivalent to UQC. 
Therefore, two noncommuting local controls -- that need not even act 
on the same qubit~\cite{Wang++:16} -- represent the minimal control resource required for UQC in qubit arrays with 
always-on Heisenberg-type interaction between adjacent qubits. This opens up the possibility of UQC through continuous-wave control, which enables single-shot realizations of quantum gates and multiqubit states.

The fact that two local controls are already sufficient for complete controllability of a Heisenberg-coupled $N$-qubit array begs another interesting question -- namely, whether an even smaller control resource (i.e., a single local control) permits some semblance of controllability in such an array. Indeed, it has already been shown
that a single local control in such an array guarantees subspace controllability~\cite{Albertini+DAlessandro:25} on an arbitrary subspace of the total Hilbert space that is characterized by a fixed number $a$ of excitations~\cite{Wang++:16}. This also implies state-to-state controllability 
on each of those $N+1$ subspaces. In other words, for an arbitrary choice of initial- and final states with the same number of excitations (i.e., the same Hamming weight) it is possible in principle to find control fields such that the system dynamics allow an evolution of the system from the given initial- to the desired final state. 

Here, we harness the subspace controllability of Heisenberg-coupled qubit array with a single local control in the form of an analog (pulse-level) 
state-engineering scheme.
Given that the Dicke state with the excitation number $a$ is the equal-weight linear combination of all possible product states with the same number of excitations, this Dicke state is reachable in the 
system at hand provided that one starts from a generic product state with the Hamming weight $a$. In other words, starting from an initial product state with $a$ excitations at $t=0$, the $a$-excitation Dicke state can be dynamically generated at a later time $t=T$.
To find the appropriate time-dependence of the control field -- acting in the $z$ direction on a single actuator qubit -- that allows one to steer the system towards the desired Dicke state as rapidly as possible, i.e., in the shortest possible time $T$, the toolbox of quantum optimal control is employed here.

The control scheme utilized here, 
which is based on the dCRAB algorithm~\cite{rach_dressing_2015},
relies on a smoothly-varying control 
field expanded over a truncated random Fourier basis of functions. The global maxima of the relevant figure of merit (target-state fidelity) as a function of the expansion coefficients of the control field acting in the $z$ direction are obtained by combining the Nelder-Mead simplex method for local optimization~\cite{NRcBook} and the 
multistart-based clustering algorithm that facilitates the search for global extrema~\cite{TornZilinskasBook}. In this manner, both the shortest possible times required 
for high-fidelity realizations of $W$ ($a=1$) and genuine Dicke states ($a\geq 2$), as well as the corresponding optimal control fields, are obtained in arrays with up to $N=9$ qubits. Based on our numerical results, we establish power-law scalings with $N$ of the shortest possible state-preparation times for $W$ and $a=2$ Dicke states. Finally, we demonstrate the robustness of our state-engineering scheme against small control-field distortions from the optimal pulse shape, a small leakage of this field away from the actuator qubit, and its small misalignment (tilt) from the nominal $z$ direction.

The remainder of this paper is structured as follows. The definition and basic properties of Dicke states are reviewed in Sec.~\ref{DickeStatesReview}. In  Sec.~\ref{ControlHeisenberg} the main 
Lie-algebraic results pertaining
to the controllability of spin-$1/2$ chains 
with nearest-neighbor Heisenberg-type coupling are recapitulated, while Sec.~\ref{SubspControlHeisenberg} covers in detail the 
subspace controllability of the same class of systems. Section~\ref{DickeDynamGener} frames the problem at hand as a quantum-control problem, as well as describing the methodology for finding optimal control fields. The main results of this work are presented for an idealized system in Sec.~\ref{IdealSysResults},
while in Sec.~\ref{SchemeRobustness} we discuss the effects of various real-world imperfections.
Finally, the paper is rounded out in Sec.~\ref{SummConcl}, which also contains a short summary of its principal findings.

\section{Dicke states: an overview}
\label{DickeStatesReview}
To set the stage for further considerations, the definition, basic properties, and concrete 
examples of Dicke states are provided below.

An $a$-excitation product state of an $N$-qubit system can be parameterized in the form
$\ket{\{n_1,\ldots,n_a\}}$, where $n_1,\ldots,n_a$ enumerate the qubits that are 
in the logical state $|1\rangle$ and the remaining qubits are in the state $|0\rangle$. 
The $N$-qubit Dicke state with $a$ excitations is given by the equal-weight superposition
of all the product states $\ket{\{n_1,\ldots,n_a\}}$, i.e.,
\begin{equation}\label{DickeStateDef1}
\ket{D^{N}_{a}}= \uber{N}{a}^{-1/2}\sum_{n_1<\ldots<n_a}^N
\ket{\{n_1,\ldots,n_a\}} \:.
\end{equation}
In other words, the state $\ket{D^{N}_{a}}$ is the equal-weight
linear combination of the $\uber{N}{a}$ product states corresponding to the bit strings 
of Hamming weight $a$. 

In the special, single-excitation ($a=1$) case 
Dicke states coincide with $W$ states, i.e.,
$\ket{D^{N}_{1}}\equiv\ket{W_{N}}$, where
\begin{equation} \label{WstateDef}
\ket{W_{N}} = \frac{1}{\sqrt{N}}\:
(|10\ldots 0\rangle + |010\ldots 0\rangle 
+ \ldots + |0\ldots 01\rangle)
\end{equation}
is the $N$-qubit $W$ state. $W$ states constitute one of the most important classes of 
maximally-entangled multiqubit 
states~\cite{Haase+:22}. Being more robust 
to particle loss than any other class of multiqubit states, they hold promise for 
a multitude of quantum-technology applications; as a result, their realizations are 
proposed in all currently investigated physical platforms for QC~\cite{StojanovicPRL:20,
Stojanovic:21,Zheng++:22,Zhang+:23,Srivastava+:25}. 

Instead of the notation used in the above definition of Dicke states [cf. Eq.~\eqref{DickeStateDef1}], 
for small $N$ it is more convenient to use a simpler notation that involves 
bit strings. For example, the two-excitation ($a=2$) Dicke state $|D^3_2\rangle$ in a 
system of $N=3$ qubits, and its counterpart $|D^4_2\rangle$ in a $4$-qubit system, are 
given by
\begin{eqnarray} \label{TwoExcDicke}
|D^3_2\rangle &=& \frac{1}{\sqrt{3}}\:(|110\rangle
+|101\rangle+|011\rangle) \:,\\
|D^4_2\rangle &=& \frac{1}{\sqrt{6}}\:(|1100\rangle
+|1010\rangle+|1001\rangle \nonumber\\
& &+ |0110\rangle + |0101\rangle + |0011\rangle) \:.\nonumber
\end{eqnarray}

It is important to note that $N$-qubit Dicke states are invariant under arbitrary permutations 
of qubits. This property is manifest in yet another useful representation of Dicke states, 
equivalent to Eq.~\eqref{DickeStateDef1}, which is given by
\begin{equation}\label{DickeStateDef2}
\ket{D^{N}_{a}}= \uber{N}{a}^{-1/2}\sum_{P\in S_N}
P\{ \ket{1}^{\otimes a} \ket{0}^{\otimes 
(N-a)} \} \:,
\end{equation}
where the sum on the RHS of the last equation runs over all permutations $P$ of the set $\{1,2,\ldots,N\}$, i.e., all 
elements of the symmetric group $S_N$. 
Moreover, Dicke states $\ket{D^{N}_{a}}$ 
($a=0,\ldots,N-1$) form a basis 
of the permutationally-invariant subspace 
(the symmetric sector) 
of the $2^N$-dimensional $N$-qubit Hilbert 
space; the fact that this last subspace has the dimension $N+1$, which grows only linearly -- rather than exponentially -- with 
the number of qubits, explains its widespread use in quantum-information processing problems involving fully permutationally-invariant systems~\cite{Stojanovic+Nauth:22,
Loetstedt+Yamanouchi:25,
Haase+:21,Nauth+Stojanovic:22,
Sharma+Bhosale:26}. 

Another property of Dicke states that is of interest for the present work is that states $\ket{D^{N}_{a}}$ and $\ket{D^{N}_{N-a}}$ are equivalent up to local change of basis $|0\rangle \rightleftarrows|1\rangle$; therefore, these states have the same entanglement properties.
In particular, $W$ states $|W_{N}\rangle\equiv \ket{D^{N}_{1}}$ and Dicke states $\ket{D^{N}_{N-1}}$ are an example of such pairs of states.

Dicke states exemplify quantum states with {\em genuine multipartite entanglement} --
i.e., states that are entangled across every bipartition (a splitting of the original multipartite system into two subsystems), but which cannot be 
expressed as statistical mixtures 
of states that only feature bipartite entanglement, without any multipartite entanglement~\cite{Toth:12}. In particular, $N$-qubit Dicke states with $N/2$ excitations (for even $N$)
$\ket{D^{N}_{N/2}}$ are known to have, for large $N$, the smallest possible overlap with states that feature no genuine multipartite entanglement~\cite{Toth:07}.

\section{Controllability / reachability for Heisenberg spin-$1/2$ chains}\label{ControlHeisenberg}
In what follows, a survey of the principal 
Lie-algebraic controllability results pertaining to Heisenberg-coupled qubit arrays is given. 
The general Lie-algebraic framework of quantum 
control is first reviewed (Sec.~\ref{basics}).
This is followed by a general introduction 
into the concept of local control (Sec.~\ref{Loc_Control}). Basic results 
pertaining to minimal control resources 
required for complete controllability 
of Heisenberg-coupled arrays are then discussed 
(Sec.~\ref{HeisenbergControl}).

Before embarking on discussions of general controllability and reachability criteria, as well as their applications to interacting spin-$1/2$ chains (qubit arrays), it is pertinent to introduce the notation that will be used in what follows for writing Hamiltonians of such systems. In particular, the single-qubit Pauli operators will be denoted by $X$, $Y$, and $Z$; using the standard computational basis, these operators can be expressed as 
$X\equiv |0\rangle\langle 1| 
+|1\rangle\langle 0|$, 
$Y\equiv i(|1\rangle\langle 0| 
- |0\rangle\langle 1|)$, 
and $Z\equiv |0\rangle\langle 
0| - |1\rangle\langle 1|$, while
$\mathbbm{1}_2 = |0\rangle\langle 0| 
+|1\rangle\langle 1|$ is the single-qubit identity operator~\cite{NielsenChuangBook}. 
The counterparts of the operators $X$, $Y$, 
and $Z$ that act in the $N$-qubit Hilbert space 
$\mathcal{H}\equiv(\mathbbm{C}^2)^{\otimes N}$ (the tensor product of single-qubit Hilbert spaces $\mathbbm{C}^2$) will be denoted by $X_n$, $Y_n$, and $Z_n$, respectively,
in the following. These last operators are jointly 
defined by the tensor-product relation
\begin{equation} \label{defineXYZ_n}
\mathbf{X}_n \equiv \mathbbm{1}_2\otimes\ldots\otimes\mathbbm{1}_2\otimes 
\underbrace{\mathbf{X}}_n\otimes\mathbbm{1}_2\otimes\ldots\otimes\mathbbm{1}_2  \:,
\end{equation}
where $\mathbf{X}\equiv(X, Y, Z)^{\textrm{T}}$ is the vector of single-qubit Pauli operators and 
$\mathbf{X}_n\equiv(X_n, Y_n, Z_n)^{\textrm{T}}$ ($n=1,\ldots,N$) that of their extension to the $N$-qubit Hilbert space. 

\subsection{Lie-algebraic controllability and reachability criteria for finite-dimensional quantum systems}\label{basics}
Let us consider a quantum system, with 
the $d$-dimensional Hilbert space 
$\mathcal{H}$. Assume that $H_0$ is the 
time-independent intrinsic (drift) Hamiltonian of this system and that the
latter is acted upon by external control fields $f_j(t)$ ($j=1,\ldots,p$). These fields couple to certain degrees 
of freedom of the system, which are represented by Hermitian 
operators $H_j$ (sometimes referred to as the drive Hamiltonians). The total Hamiltonian of the system is then given by
\begin{equation}\label{General_Hamiltonian}
H(t)=H_0+\sum_{j=1}^{p}f_j(t)H_j \:.
\end{equation}
The time-evolution operator $U(t)$ of the system satisfies 
the dynamical (Schr\"{o}dinger) equation ($\hbar=1$)
\begin{equation}\label{time_evolution}
\frac{dU}{dt} = -i[H_0 +\sum_{j=1}^{p}f_j(t)H_j]\:U (t)\:, 
\end{equation}
with the initial condition $U(t=0) =\mathbbm{1}_{d}$, where $\mathbbm{1}_{d}$ is the identity operator on the Hilbert space 
$\mathcal{H}$~\cite{D'AlessandroBook}. The objective of a 
generic quantum-control problem is to determine a time $T > 0$ and time-dependent 
controls $f_j(t)\in \mathbbm{R}$ such that a desired 
unitary $U_{\textrm{target}}$ is reached at $t=T$;
in other words, $U(t=T)=U_{\textrm{target}}$. 
In particular, the system is said to be completely (operator) controllable provided that its dynamics, governed by $H(t)$,
gives rise -- with a certain choice of the fields 
$f_j(t)$ -- to an arbitrary unitary on its Hilbert space. 
Rephrasing, complete controllability assumes that the 
reachable set of the system -- i.e., the set of unitaries achievable by varying the controls -- is given by  
the Lie group U$(d)$ or SU$(d)$~\cite{D'AlessandroBook}.

The criteria of controllability for quantum systems are framed using Lie-algebraic concepts~\cite{Jurdjevic+Sussmann:72,
Huang+Tarn+Clark:83,Ramakrishna+Rabitz:96,
D'AlessandroBook}, among which
that of the dynamical Lie algebra (DLA)
of the system plays the principal role~\cite{D'AlessandroBook}. For a system governed by
the Hamiltonian in Eq.~\eqref{General_Hamiltonian}, the DLA
$\mathcal{L}$ is generated by the skew-Hermitian counterparts of the operators $H_k$. i.e., by the operators $\{-iH_k|k=0,\ldots,p\}$. The necessary and sufficient condition for complete controllability (the Lie-algebraic rank condition)~\cite{D'AlessandroBook} is that $\mathcal{L}$ be isomorphic to the Lie algebra 
$\mathfrak{u}(d)$ of 
skew-Hermitian $d\times d$ matrices, or its counterpart $\mathfrak{su}(d)$ that corresponds to  traceless skew-Hermitian ones~\cite{PfeiferBook}. This general result constitutes an existence theorem that guarantees
that any unitary on the Hilbert space of the system is reachable for appropriately chosen control fields. Another,
completely separate, question is how to find the
actual time dependence of control fields that allows
one to realize a desired unitary. This is typically done
by taking into account various additional constraints -- e.g., the one pertaining to the total duration of the control protocol. 

\subsection{Local (quantum) control: general concept and its application to qubit arrays} \label{Loc_Control}
The principal control-related issue pertaining to interacting quantum systems is whether a given system can be fully- or,
at least, partially controlled by solely acting on its subsystem. 
This is the idea of the {\em local-control} approach, which assumes that only a small subsystem of the original system is subject to control fields. The actual choice of the subsystem crucially depends on the type of interaction in the considered quantum system.

Let us consider a composite system $S=C\cup \bar{C}$, described by the Hamiltonian 
\begin{equation} \label{HamiltLocal}
H(t)=H_S+\sum_j\:f^{C}_j(t) H^{C}_j \:,   
\end{equation}
where $H_S$ is the drift part (acting on the whole system $S$), while $H^{C}_j$ are local Hamiltonians (acting only on the subsystem $C$) and $f^{C}_j(t)$ the corresponding time-dependent control fields (for an illustration, see Fig.~\ref{LocalCtrlIllust:fig1}). Assuming, for simplicity, that $-iH_j^{C}$'s  generate the Lie algebra $\mathcal{L}(C)$ on $C$, the system $S$ 
is completely controllable if and only 
if $-iH_S$ and $-iH^{C}_j$ generate $\mathcal{L}(S)$, i.e., 
\begin{equation}
\langle iH_S,\mathcal{L}(C)\rangle=\mathcal{L}(S) \:,
\end{equation}
where $\langle A,B\rangle$ stands for the algebraic closure of the operator sets $A$ and $B$~\cite{PfeiferBook}. 
Therefore, any unitary on $S$ can be enacted through control of its subsystem $C$ if and only if every element of $\mathcal{L}(S)$ can be obtained either as a linear combination 
of $-iH_S$, $-iH^{C}_j$ (i.e., skew-Hermitian counterparts of the drift and drive Hamiltonians), or their nested commutators.

\begin{figure}[b!]
\includegraphics[clip,width=0.95\columnwidth]{GenDickeSubsControl_Fig1.pdf}
\caption{\label{LocalCtrlIllust:fig1}Pictorial illustration of 
the concept of local control. System $S$ is described 
by the drift Hamiltonian $H_S$. The system is subject 
to external control fields $f_j^{C}(t)$, which couple only to its degrees of freedom that belong to the subsystem $C$; 
these degrees of freedom are described by the local Hamiltonians $H_j^{C}$. The total Hamiltonian of the system is 
then the one given by Eq.~\eqref{HamiltLocal}.}
\end{figure}

Qubit arrays constitute a prototypical class of systems in which local-control approach may be 
advantageous~\cite{LocalPlastina}. 
In keeping with the above Lie-algebraic criteria (cf. Sec.~\ref{basics}), complete controllability of an $N$-qubit array 
is that its corresponding DLA be isomorphic with either $\mathfrak{u}(2^N)$ or $\mathfrak{su}(2^N)$. The conventional approach to control in a qubit array requires control fields to act on each qubit in the array. Along with a drift Hamiltonian, which describes qubit-qubit interactions that make it possible to realize 
entangling two-qubit gates, this -- in 
principle -- permits the realization of an 
arbitrary (multi-qubit) gate. In contrast to this scenario, in the local-control case 
control fields act only on a handful
of {\em actuator} qubits -- in the extreme case,
a single qubit. The specific choice of actuator 
qubits in a given system should ideally be one 
guaranteeing complete controllability, as the latter renders UQC possible~\cite{D'AlessandroBook}. 

The usefulness of local control in qubit arrays cannot be overstated, especially
in the context of solid-state qubits. In principle, global-control approaches to qubit arrays constitute a promising pathway to scalable QC~\cite{Veldhorst+:15}. In particular, 
a continuous-wave global field permits decoupling of the qubits from background noise. In the case of solid-state qubits this global-control approach is, however, compromised by the unavoidable variability in the parameters of individual qubits in the array 
(e.g. in the qubit resonance frequency). 
Therefore, applications of global-control strategies in solid-state QC platforms usually require additional nontrivial 
steps~\cite{Hansen+:21,George+:25,Aiudi+:26}, in systems at the noisy intermediate scale quantum level and beyond~\cite{PreskillNISQ:18}. 

\subsection{Complete controllability of Heisenberg-coupled spin-$1/2$ chains (qubit arrays): Principal results}\label{HeisenbergControl}
The issue of identifying the minimal control resources required for complete controllability of various interacting spin-$1/2$ models (Ising, $XY$, Heisenberg, etc.) was studied extensively in the past~\cite{Schirmer++:08,Wang++:16}. For each of 
these models, the 
smallest subsystem was sought that -- when acted 
upon by Zeeman-type external controls -- renders 
the entire chain completely controllable. As it turned out, only for Heisenberg-type coupling these investigations yielded nontrivial results.

The most general controllability-related result for spin-$1/2$ chains (qubit arrays) with Heisenberg-type interaction was proven based on a method that makes use of the Hilbert-space decomposition into a tensor product of minimal invariant subspaces~\cite{Wang++:16}. This result
asserts that the existence of two mutually noncommuting 
local controls -- which need not act on the 
same spin (qubit) -- guarantees complete controllability 
of the chain; importantly, this last result holds even 
when for fully anisotropic $XYZ$ coupling 
case~\cite{Wang++:16}, i.e., for the drift Hamiltonian
\begin{equation}\label{H_XYZ}
H_{XYZ}=\sum_{n=1}^{N-1}\:\left(J_x 
X_{n}X_{n+1}+J_y Y_{n}Y_{n+1} +
J_z Z_{n}Z_{n+1}\right) \:,
\end{equation}
where $J_x$, $J_y$, and $J_y$ are the three coupling 
strengths. While the last controllability-related result does not require the two noncommuting controls to act 
on the same qubit, the simplest scenario to which
this theorem applies is the one in which the first
qubit in the array is subject to a Zeeman-type field
control field $\mathbf{B}_1 (t)$ with two nonzero components. Assuming, for definiteness, that these components are $x$ and $y$, i.e., that $\mathbf{B}_1 (t)\equiv[\:B_{1x}(t),B_{1y}(t),0\:]^{\textrm{T}}$, the corresponding control Hamiltonian is given by
\begin{equation}\label{controlham}
H_c(t)=B_{1x}(t) X_{1}+B_{1y}(t) Y_{1} \:.
\end{equation} 

The stated general controllability result, pertaining to the drift Hamiltonian $H_{XYZ}$ [cf. Eq.~\ref{H_XYZ}], has two special cases that had 
been proven long before the general one. The first 
of these special cases is that of the $XXZ$ drift Hamiltonian
\begin{equation}\label{H_XXZ}
H_{XXZ}=J\sum_{n=1}^{N-1}\:\left(X_{n}X_{n+1}+Y_{n}
Y_{n+1}+\Delta Z_{n}Z_{n+1}\right) \:,
\end{equation}
where $J_x=J_y\equiv J$ and $J_z\equiv J\Delta$, with 
$\Delta$ being the anisotropy parameter.
The second special case corresponds to the fully isotropic Heisenberg Hamiltonian (i.e., the one with 
$J_x=J_y=J_z \equiv J$)
\begin{equation}\label{H_XXX}
H_{XXX}=J\sum_{n=1}^{N-1}\:\left(X_{n}X_{n+1}
+ Y_{n}Y_{n+1} + Z_{n}Z_{n+1}\right) \:,
\end{equation}
which happens to be of most relevance for 
applications in realistic qubit arrays~\cite{George+:25}. 

\begin{figure}[t!]
\includegraphics[width=0.95\columnwidth]{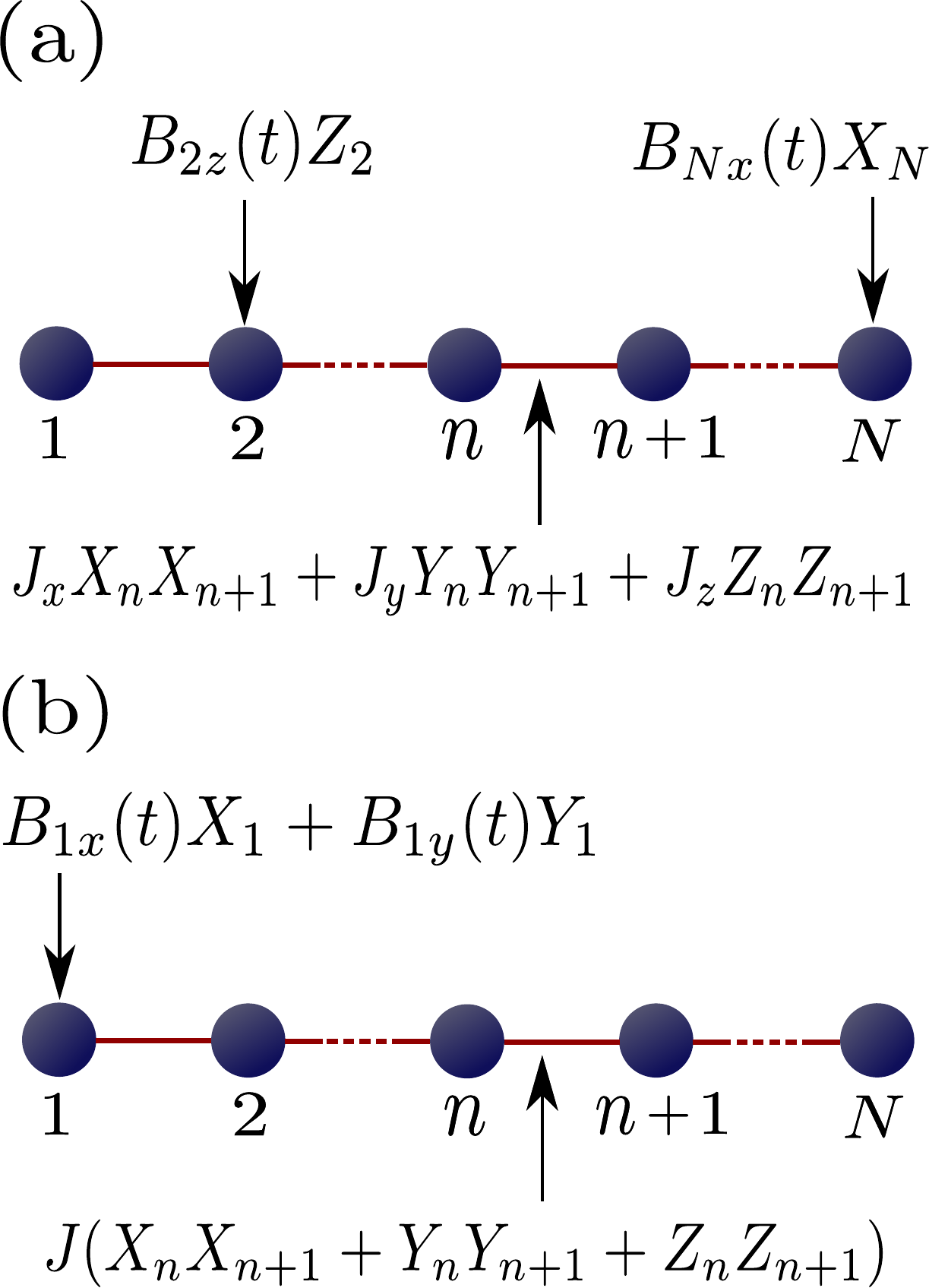}
\caption{\label{CompCtrlHeisenberg:fig2}Schematic illustration of possible complete-controllability scenarios in Heisenberg-coupled $N$-qubit arrays with a local control: (a) In an array with the fully anisotropic Heisenberg interaction a local $Z$ control is applied to qubit $2$ and local $X$ control to qubit $N$, (b) In an array with the isotropic Heisenberg interaction local $X$- and $Y$ controls are both applied to qubit $1$.}
\end{figure}

Two extreme scenarios that permit complete controllability of Heisenberg-coupled qubit arrays are illustrated in Fig.~
\ref{CompCtrlHeisenberg:fig2}. An example of a qubit array with the fully anisotropic Heisenberg interaction [cf. Eq.~\eqref{H_XYZ}] and two noncommuting local Zeeman-type controls applied to different qubits -- more precisely, one interior qubit (qubit $2$) and one of the end qubits (qubit $N$) -- is shown Fig.~\ref{CompCtrlHeisenberg:fig2}(a). On the other hand, an example of a qubit array with the isotropic Heisenberg interaction [cf. Eq.~\eqref{H_XXX}] and two noncommuting local Zeeman-type controls, both applied to same end qubit (qubit $1$) is depicted in Fig.~\ref{CompCtrlHeisenberg:fig2}(b).

For each of the Hamiltonians in Eqs.~\eqref{H_XYZ}-\eqref{H_XXX}, complete controllability of the underlying system can be proven by demonstrating that its corresponding DLA, generated by the skew-Hermitian 
counterparts of the drift Hamiltonian [i.e., $-iH_{XYZ}$,
$-iH_{XXZ}$, $-iH_{XXX}$, for the drift Hamiltonians
in Eqs.~\eqref{H_XYZ}-\eqref{H_XXX}, respectively] 
and the two single-spin(qubit) Pauli operators describing noncommuting controls [e.g., $-iX_1$ and $-iY_1$ in 
the example of the control Hamiltonian in Eq.~\eqref{controlham}], has the dimension equal to $d^2-1$ (note that each of the above drift Hamiltonians, as well as the control Pauli operators, is traceless) with $d\equiv 2^{N}$ 
being the dimension of the Hilbert space of the system. This implies that the relevant DLA is isomorphic with the Lie algebra $\mathfrak{su}(d)$~\cite{PfeiferBook}.

\section{Subspace controllability 
of Heisenberg spin-$1/2$ chains}
\label{SubspControlHeisenberg}
The crucial control-related concept for the remainder of this work,  providing the theoretical underpinning for the optimal-control-based generation of Dicke states, is that of subspace controllability. General aspects of this concept are discussed in Sec.~\ref{SubspControlGeneral} below. The discussion in Sec.~\ref{SubspControl_XXZandXYZ} specializes to subspace controllability of
$XXZ$- and $XYZ$-type Heisenberg spin-$1/2$ chains.
Finally, in Sec.~\ref{ObstructControllability} we point out a special case of the latter systems in which
subspace controllability is partially obstructed due to an additional 
symmetry.

\subsection{Subspace controllability: general aspects}
\label{SubspControlGeneral}
Among different situations that can be encountered when discussing controllability of finite-dimensional quantum systems, subspace controllability refers to the one in which the underlying Hilbert space can be split into the 
direct sum of invariant subspaces 
$\mathcal{H}_{\alpha}$ and, 
on each of these subspaces,
it is possible to generate any arbitrary unitary operation using appropriately chosen control functions~\cite{Wang++:16} -- in other words, the DLA $\mathcal{L}$ of the system can be written in the form of the direct sum 
$\mathcal{L}=\oplus \mathcal{L}_{\alpha}$, where its reduction 
$\mathcal{L}_{\alpha}$ to the 
invariant subspace $\mathcal{H}_{\alpha}$ is isomorphic with $\mathfrak{su}(\textrm{dim}\mathcal{H}_{\alpha})$.
This also implies state-to-state controllability for any choice of initial- and final states that belong to this same subspace. 

Familiar examples of subspace controllability are
furnished by permutationally-symmetric networks of qubits~\cite{Chen+:20} -- e.g., that of all-to-all Ising($zz$)-coupled qubits subject to transverse ($x$ and $y$) global control fields~\cite{Stojanovic+Nauth:22,
Stojanovic+Nauth:23,Evangelakos+:24,Evangelakos+:25}, where the relevant subspace is the $(N+1)$-dimensional  permutationally-invariant subspace (symmetric sector) of the total $N$-qubit Hilbert space (cf. Sec.~\ref{DickeStatesReview})~\cite{Zhou+:26}. Another example is that of $XY$-coupled spin-$1/2$ chain (qubit array) with a single local control [cf. 
Sec.~\ref{SubspControl_XXZandXYZ} below], where subspace controllability holds only for the single-excitation 
subspace~\cite{Wang++:16}. 

The existence of controllable subspaces is common in cases where the system under consideration possesses dynamical symmetries~\cite{Albertini+DAlessandro:25}. In the context of quantum control [recall Sec.~\ref{basics}], with the drift Hamiltonian $H_0$ and $p$ external controls described by Hamiltonians $H_j$ ($j=1,\ldots,p$), any Hermitian operator $S$ that 
is not a multiple of the identity and commutes with both $H_0$ and $H_1,\ldots,H_p$ (i.e., $[H_0 ,S]=[H_j ,S] = 0$) is referred to as the symmetry operator of the system. It follows directly that the operators $H_0$, $H_j$  ($j=1,\ldots,p$), and $S$ can be simultaneously diagonalized. 

\subsection{Subspace controllability of $XXZ$- and $XYZ$-type 
Heisenberg-coupled spin-$1/2$ chains} \label{SubspControl_XXZandXYZ}
The existing studies of subspace controllability of Heisenberg-coupled spin-$1/2$ chains showed that for this class of models 
subspace controllability
depends on whether 
the (two-local) interactions in 
the two directions perpendicular to the control direction are equal or not, i.e., whether the 
spin-$1/2$ chain is of $XXZ$
type -- including the fully
isotropic $XXX$ case -- or anisotropic $XYZ$ type~\cite{Wang++:16}. Due to the significantly higher degree of symmetry 
in the $XXZ$ case, the character of dynamical symmetries -- accordingly, that of invariant subspaces as well -- is different 
in these two cases.

In the case of  
$XXZ$-type Hamiltonians [cf. Eq.~\eqref{H_XXZ}] -- including the special case $\Delta=1$ [i.e., the isotropic ($XXX$) Heisenberg 
Hamiltonian in Eq.~\eqref{H_XXX}] -- it is pertinent to introduce the excitation operator
\begin{equation}\label{exc_oper}
S_{\textrm{exc}} =\frac{1}{2}\:\sum_{n=1}^{N}
(\mathbbm{1}_2 + Z_n)  \:.
\end{equation}
It is worthwhile noting that this operator
can be recast as $S_{\textrm{exc}} =
S_z+N/2$, where  
\begin{equation}\label{defSz}
S_z = \frac{1}{2}\:\sum_{n=1}^{N} Z_n 
\end{equation}
is the $z$ projection of the collective-spin operator of the system.
The excitation operator has $N + 1$ distinct eigenvalues $a= 0,1,\ldots,N$. Therefore, the Hilbert space of the 
system can be decomposed as a direct 
sum of its eigensubspaces, i.e., 
$\mathcal{H}= \bigoplus_{a=0}^{N}\mathcal{H}_a$.
In particular, the eigensubspace $\mathcal{H}_a$ is generated by the 
states corresponding to bit-strings 
with exactly $a$ occurrences of $1$, i.e., by the Hamming-weight-$a$ states.

It is straightforward to verify that $[H_{XXZ},S_{\textrm{exc}}]=0$. Provided that the relevant 
control Hamiltonians 
$H_j$ contain only Pauli-$Z$ operators, then $S_{\textrm{exc}}$
commutes with all of them and defines a symmetry,
which is referred to as the excitation symmetry. As a result, both the Hamiltonian $H_{XXZ}$ and control Hamiltonians can be block-diagonalized on each of the $N + 1$ 
eigensubspaces $\mathcal{H}_a$ of $\mathcal{H}$, which in that case also represent the invariant subspaces of the system. It is straightforward to see that for an 
$N$-qubit system, the dimension of $\mathcal{H}_a$
is given by 
\begin{equation}
\textrm{dim}\:\mathcal{H}_a = \uber{N}{a} = \frac{N!}{a!(N-a)!} \:.
\end{equation}
In particular, the largest of these invariant subspaces is the one that for $N$ even corresponds to $a=N/2$ (in the case of odd $N$ it corresponds to 
$a=\lfloor N/2 \rfloor$); the asymptotic dimension of this subspace for large $N$ grows exponentially with $N$; namely, by 
making use of the Stirling formula $N!\approx \sqrt{2\pi N}(N/e)^N$ (for large $N$) one finds that $\textrm{dim}\:\mathcal{H}_{N/2}\sim 2^N/\sqrt{\pi/2N}$ for large $N$.

As already proven in \cite{Wang++:16}, an $XXZ$ 
spin-$1/2$ chain of length $N$ with a 
single local $Z$ control on an end spin
is controllable on each of the $N + 1$ 
invariant excitation subspaces $\mathcal{H}_0,
\ldots,\mathcal{H}_N$. It is important
to stress that this last result also holds for
the special case of an isotropic Heisenberg 
$XXX$ chain. However, in the latter case, owing to the absence of preferred spatial direction
(i.e., the symmetry between the $x$, $y$, and $z$ directions), this result applies for a local control in any of the three directions.

\begin{figure}[t!]
\includegraphics[width=0.95\columnwidth]{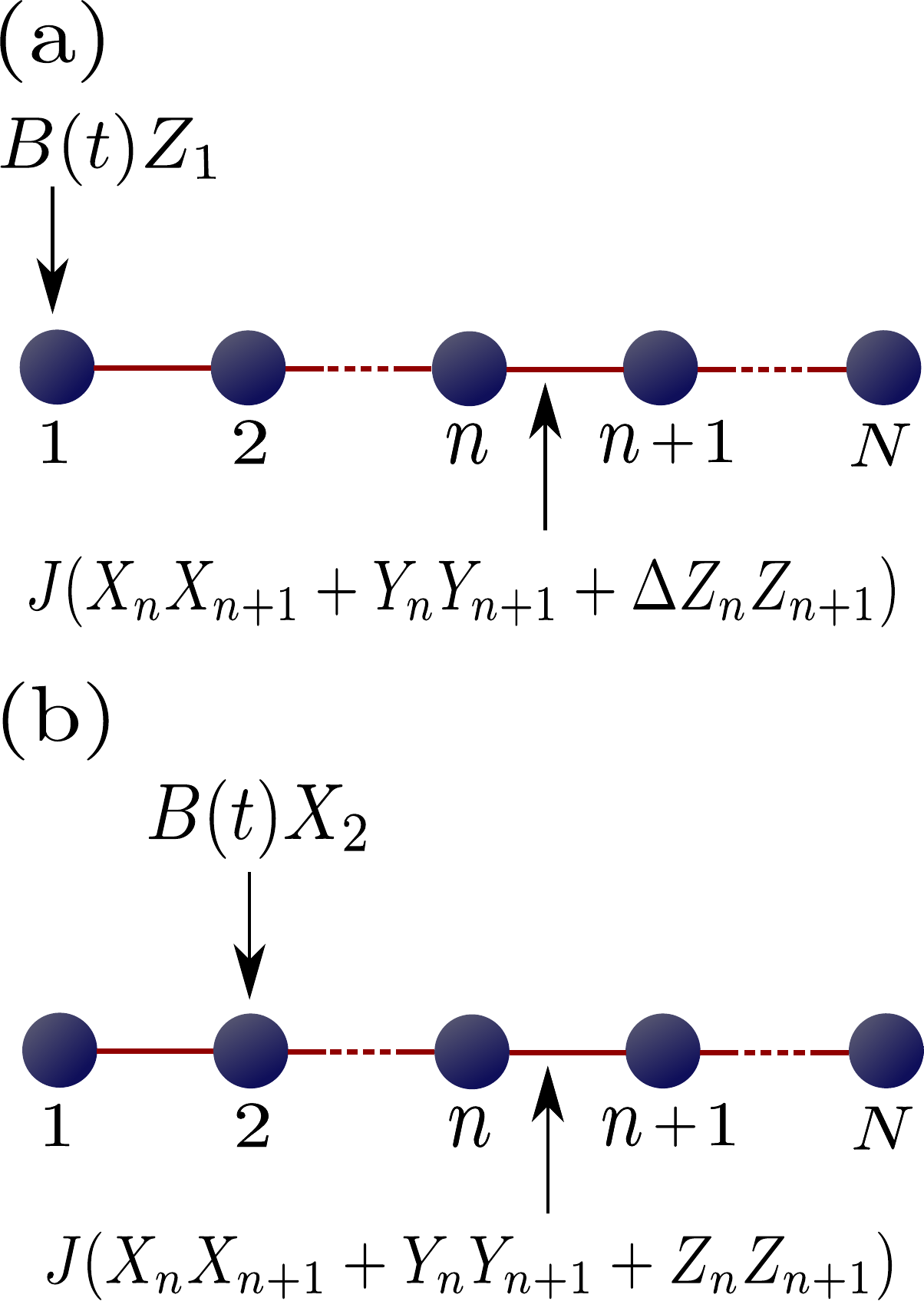}
\caption{\label{SubspControlXXZXXX:fig3}Schematic illustration of possible subspace-controllability scenarios in Heisenberg-coupled 
$N$-qubit arrays with a local control effected through a unidirectional time-dependent magnetic field of magnitude $B(t)$:
(a) In an array with the $XXZ$-type Heisenberg interaction a local $Z$ control is applied to qubit $1$, (b) In an array with the isotropic Heisenberg interaction a local $X$ control is applied to 
qubit $2$. }
\end{figure}

Two possible subspace-controllability scenarios in spin-$1/2$ chains (qubit arrays) with 
$XXZ$-type Heisenberg interaction are illustrated in Fig.~\ref{SubspControlXXZXXX:fig3}.
While Fig.~\ref{SubspControlXXZXXX:fig3}(a) depicts the most general $XXZ$-interaction case -- where a single local control ought to be applied in the $z$ direction -- Fig.~\ref{SubspControlXXZXXX:fig3}(b) illustrates the special case of isotropic Heisenberg interaction, where a control in the $x$ direction is equally permissible as one in the $z$ direction.

By contrast to the $XXZ$ case, for the $XYZ$ model [cf. Eq.~\eqref{H_XYZ}] the excitation symmetry [represented by the operator $S_{\textrm{exc}}$
in Eq.~\eqref{exc_oper}] -- is not one of the 
dynamical symmetries. However, this model does have the $Z$-parity symmetry, represented by the operator
$S_P = Z_1 Z_2\ldots Z_N$; $H_{XYZ}$ has two $2^{N-1}$-dimensional invariant subspaces $\mathcal{H}_1$ and 
$\mathcal{H}_{-1}$, which correspond to the eigenvalues 
$\pm 1$ of $S_P$, respectively. Therefore, compared to the $XXZ$ case, the number of invariant subspaces in the $XYZ$ chain is reduced from $N+1$ to $2$ as a consequence
of symmetry breaking between the $x$ and $y$ directions. 
Similarly to the $XXZ$ case, it was proven that an 
$XYZ$ spin-$1/2$ chain of length $N$ with a single
local $Z$ control on an end spin, the system is indeed controllable on each of the two invariant subspaces 
$\mathcal{H}_1$ and $\mathcal{H}_{-1}$~\cite{Wang++:16}.

The above results on
subspace controllability of $XXZ$ and $XYZ$ Heisenberg spin-$1/2$ chains with a single local control can succinctly be expressed in terms of their respective DLAs.
In the $XXZ$ case, the DLA of an $N$-spin chain is given by the direct sum $\mathcal{L}=\oplus_{a=0}^{N-1}\mathcal{L}_a$, 
where $\mathcal{L}_a \cong \mathfrak{su}(\textrm{dim}\mathcal{H}_a)$ is the reduction of $\mathcal{L}$ to the $a$-excitation invariant subspace $\mathcal{H}_a$. On the other hand, in the $XYZ$ case
the DLA of an $N$-spin chain can be written in the form $\mathcal{L}=\mathcal{L}_{+}\oplus\mathcal{L}_{-}$, where $\mathcal{L}_{\pm}\cong\mathfrak{su}(2^{N-1})$
are the respective 
reductions of the DLA 
$\mathcal{L}$ of the spin chain 
to the $2^{N-1}$-dimensional 
invariant subspaces 
$\mathcal{H}_{\pm 1}$.

For the sake of completeness, it is worthwhile mentioning that in the absence of the $ZZ$ 
coupling [i.e., in the special case $J_z=0$ of 
Eq.~\eqref{H_XYZ}]-- i.e., for the $XY$-type Hamiltonians (including its special case 
$J_x=J_y\equiv J$, which is known as the $XX$ Hamiltonian; such Hamiltonians are quite 
common in the realm of superconducting qubits, 
where coupling may go 
beyond nearest-neighbor qubits) -- the subspace controllability holds 
only in the 
single-excitation subspace ($a=1$). It is also 
useful to mention that an interesting class of 
highly-entangled multiqubit states  that belong to
such a subspace is that of $W$-type (Hamming-weight-$1$) states; the most important subclass of such states are the conventional, maximally-entangled $W$ states [cf. Eq.~\eqref{WstateDef}]. Another interesting subclass of Hamming-weight-$1$ maximally-entangled multiqubit states is furnished by the generalized (twisted)
$W$ states, defined for an underlying spatially periodic (i.e., one characterized by a discrete translational symmetry) arrangement of qubits~\cite{Haase+:22}; unlike the conventional $W$ states [cf. Sec.~\ref{DickeStatesReview}], these generalized states are not permutationally symmetric.

Without significant loss of generality, the analysis of the dynamical generation of Dicke states in the remainder of this work (see Secs.~\ref{DickeDynamGener}- \ref{SchemeRobustness} below) will be carried out only in the isotropic Heisenberg-interaction case, described by the Hamiltonian $H_{XXX}$. This choice is primarily motivated by the practical relevance of isotropic Heisenberg interactions in realistic solid-state qubit arrays, 
e.g. those based on semiconductor spin qubits~\cite{George+:25}.

\subsection{Obstruction of subspace controllability: Spin-$1/2$ chains with odd length and center-spin actuator}\label{ObstructControllability}

While subspace controllability of Heisenberg-coupled spin-$1/2$ chains with a single local control holds in principle for an arbitrary choice of the actuator spin, it is pertinent to underscore at this point that a special case exists in which this limited controllability is partially obstructed due 
to the presence of an additional symmetry (relative to the generic case of such a chain). This is the case of chains that contain an odd number $N$ of spins, with the center spin [i.e., the one corresponding to $n=(N+1)/2$] playing the role of the actuator. Such a chain possesses a mirror (left-right) reflection symmetry (also known as {\em parity symmetry}) with respect to the actuator;
as a result, the DLA 
$\mathcal{L}$ of the spin-$1/2$ chain decomposes into symmetry-adapted sectors instead of acting irreducibly (as its reduction $\mathcal{L}_a$) to each subspace $\mathcal{H}_a$ with a fixed excitation number $a$ $(a=0,\ldots,N-1)$. Consequently, instead of controllability on each of those subspaces, one 
finds -- at best -- controllability within smaller, symmetry-adapted sectors thereof.

To explain the aforementioned partial obstruction of subspace controllability, we first define the mirror-reflection operator $R$ through its action on an arbitrary product state of 
a spin-$1/2$ chain with odd length $N$:
\begin{equation}
R\:|\xi_1\ldots \xi_{(N+1)/2}\ldots \xi_{N}\rangle=
|\xi_{N}\ldots\xi_{(N+1)/2}\ldots\xi_1\rangle \:.
\end{equation}
Because the chain is symmetric with respect to the center spin, 
it holds that $RH_{XXX}R= H_{XXX}$. Likewise, because 
the center spin plays the role of the actuator, the control 
part $H_{C}(t)$ of the total Hamiltonian $H(t)=H_{XXX}+H_{C}(t)$ 
of the system is also invariant under reflection, i.e.,
$RH_{C}(t)R= H_{C}(t)$. Accordingly, each element of the DLA $\mathcal{L}$ also commutes with $R$.

Given that $R$ is an involution (i.e., $R^2=\mathbbm{1}$), its eigenvalues are $\pm 1$. Consequently, each excitation subspace $\mathcal{H}_a$ splits into even/odd ($\pm 1$) parity (symmetry-adapted) sectors $\mathcal{H}^{(\pm)}_a = \textrm{span}\{|\varphi^{\pm}_j\rangle\:,\: j=1,2,\ldots,\textrm{dim} \mathcal{H}^{(\pm)}_a\}$:
\begin{equation}
\mathcal{H}_a = \mathcal{H}^{(+)}_a \oplus\mathcal{H}^{(-)}_a \quad (\:a=0,\ldots,N-1\:) \:.
\end{equation}
The system dynamics cannot couple these sectors because all reachable unitaries (cf. Sec.~\ref{basics}) commute with $R$. Mathematically speaking,
the reduced DLA $\mathcal{L}_a$ of the system is 
not isomorphic to $\mathfrak{su}(\textrm{dim} \mathcal{H}_a)$, but instead to a subalgebra of $\mathfrak{su}(\textrm{dim} \mathcal{H}^{(+)}_a)\otimes \mathfrak{su}(\textrm{dim} \mathcal{H}^{(-)}_a)$. This implies {\em inter alia} that a quantum state of the odd-length spin-$1/2$ chain under consideration that belongs to one of these sectors, say 
$\mathcal{H}^{(+)}_a$, can be obtained through the control-aided dynamical evolution of the system only if its initial state $|\psi(t=0)\rangle$ belongs to that same sector.

To illustrate the above general symmetry-related considerations, it is instructive to discuss in detail the special case with $N=5$ and $a=3$, under the assumption that the third spin plays the role of the actuator.
In that case $\mathcal H_3
=\mathcal H_3^{(+)}
\oplus\mathcal H_3^{(-)}$,
where $\textrm{dim}\mathcal{H}_3=10$,
while for the two symmetry-adapted sectors we have 
$\textrm{dim}\mathcal{H}^{(+)}_3=6$ and $\textrm{dim}\mathcal{H}^{(-)}_3=4$. The basis of the parity $+1$ sector $\mathcal H_3^{(+)}$ is given by
\begin{eqnarray}
|\varphi_1^+\rangle
=\frac{|11100\rangle+|00111\rangle}{\sqrt2} \: &,& \:
|\varphi_2^+\rangle
=\frac{|11010\rangle+|01011\rangle}{\sqrt2}\:,\nonumber\\
|\varphi_3^+\rangle
=\frac{|11001\rangle+|10011\rangle}{\sqrt2}\: &,& \:
|\varphi_4^+\rangle
=\frac{|10110\rangle+|01101\rangle}{\sqrt2}\:,\nonumber\\
|\varphi_5^+\rangle
=|10101\rangle \: &,& \: |\varphi_6^+\rangle
=|01110\rangle\:,\label{BasisH3pl}
\end{eqnarray}
while its counterpart in the 
parity $-1$ sector $\mathcal H_3^{(-)}$ consists of the states
\begin{eqnarray}
|\varphi_1^-\rangle=\frac{|11100\rangle
-|00111\rangle}{\sqrt2}
\: &,& \:|\varphi_2^-\rangle
=\frac{|11010\rangle-|01011\rangle}{\sqrt2}\:,\nonumber\\
|\varphi_3^-\rangle
=\frac{|11001\rangle-|10011\rangle}{\sqrt2} \: &,& \:  |\varphi_4^-\rangle
=\frac{|10110\rangle-|01101\rangle}{\sqrt2}\:.\nonumber
\end{eqnarray}
Therefore, a parity $+1$ ($-1$) state of this $N$-spin chain can be dynamically generated in a finite time $T$ in this system only if its initial state is a 
linear combination of states $|\varphi_1^+\rangle,\ldots,
|\varphi_6^+\rangle$ (respectively,
$|\varphi_1^-\rangle,\ldots,|\varphi_4^-\rangle)$.

Instead of being isomorphic with $\mathfrak{su}(10)$, the reduction 
$\mathcal{L}_{a=3}$ of the DLA of $N=5$
Heisenberg-coupled spin-$1/2$ chains with center-spin actuator on the three-excitation subspace $\mathcal{H}_{a=3}$ is a subalgebra of $\mathfrak{su}(6)\oplus \mathfrak{su}(4)$. As stated in general above, this is a consequence of the reflection (parity) symmetry with respect to the actuator.

\section{Dynamical generation of Dicke states with a local control}\label{DickeDynamGener}
The subspace-controllability results 
reviewed in Sec.~\ref{SubspControl_XXZandXYZ} are 
utilized in the following for the dynamical
generation of Dicke states in qubit arrays
with the isotropic Heisenberg interaction
between adjacent qubits. The connection
between these subspace-controllability results and the reachability of Dicke states
is explained in Sec.~\ref{DickeStateGenSubspCtrl} below, where the Dicke-state generation is also framed as a quantum-control problem. In Sec.~\ref{MethodsOptCtrl}, various approaches to quantum optimal control are briefly surveyed, with emphasis on their potential advantages  for solving this last problem. Finally, the dCRAB algorithm -- our method of choice for designing optimal control fields in the problem at 
hand -- is briefly described in Sec.~\ref{describe_dCRAB}.

\subsection{Generation of $|D^{N}_a\rangle$ in qubit arrays with isotropic Heisenberg-type interaction}\label{DickeStateGenSubspCtrl}
As already expounded in Sec.~\ref{SubspControl_XXZandXYZ}, an array of $N$ qubits with nearest-neighbor isotropic Heisenberg interaction and a single local control acting in $x$, $y$, or $z$ direction satisfy the criteria for subspace controllability on any subspace $\mathcal{H}_a$ with the fixed number $a$ of excitations ($a=1,2,\ldots,N-1$). All such qubit arrays are described by time-dependent Hamiltonians of the form
\begin{eqnarray}
H(t) = &J&\sum_{n=1}^{N-1}\:\left(X_n
X_{n+1}+Y_{n}Y_{n+1}+Z_{n}Z_{n+1}\right) \nonumber\\
&+&\mathbf{B}(t)\cdot\mathbf{X}_{n_c} \:, \label{TotalHamIsoHeisSingleCtrl}
\end{eqnarray}
where $\mathbf{B}(t)$ stands for the unidirectional external Zeeman-type control field and $\mathbf{X}_{n_c}\equiv (X_{n_c}, Y_{n_c}, Z_{n_c})$ is the vector of Pauli operators [cf. Sec.~\ref{ControlHeisenberg}] corresponding to the actuator qubit $n=n_c$. For 
instance, $\mathbf{B}(t)=[0,0,B(t)]$ and 
$n_c =1$ correspond to the situation where qubit $1$ is acted upon by a local Pauli-$Z$ control;
similarly, $\mathbf{B}(t)=[B(t),0,0]$ and $n_c =2$ if qubit $2$ is acted upon by a Pauli-$X$ control. 

One immediate implication of the subspace controllability of qubit arrays described 
by the Hamltonians in Eq.~\eqref{TotalHamIsoHeisSingleCtrl}
is the state-to-state controllability of such
a system on each of the $N+1$ subspaces; therefore, for an initial $N$-qubit state with a certain number $a$ of excitations, any other 
$N$-qubit state with the same number of excitations (i.e., the same Hamming weight) is reachable -- through an appropriately chosen time-dependent control field -- in this system.

The established reachability of an arbitrary $N$-qubit state with a fixed number $a$ of excitations -- from any other state with the same number of excitations -- can be harnessed for the purpose of engineering Dicke states in the system under consideration. The $N$-qubit Dicke state $|D^{N}_{a}\rangle$ with $a$ excitations
($a=1,2,\ldots, N-1$), the equal-weight linear combination of all possible product states with the same number of excitations [cf. Eqs.~\eqref{DickeStateDef1} and \eqref{DickeStateDef2}], is reachable in the system at hand provided that one starts from a generic product state with the Hamming weight 
$a$; all such product states can succinctly be written in the form
$P\{\ket{1}^{\otimes a}\ket{0}^{\otimes 
(N-a)}\}$, where $P$ is an arbitrary 
permutation of the set $\{1,2,\ldots,N \}$.
Therefore, if one starts from, e.g.,  
the initial state
\begin{equation} \label{init_state}
|\psi(t=0)\rangle = \ket{1}^{\otimes a} 
\ket{0}^{\otimes (N-a)} \equiv 
|\underbrace{11\ldots 1}_a 
\underbrace{00\ldots 0}_{N-a} \rangle\:,
\end{equation}
then the above Lie-algebraic result guarantees
that an appropriate time dependence of the 
field $B(t)$ can be found such that at a later
time $T$ the Dicke state $|D^{N}_{a}\rangle$
of the $N$-qubit array is engineered, i.e.,
\begin{equation}
|\psi(t=T)\rangle = |D^{N}_{a}\rangle\:.
\end{equation}
In this manner, the state $|D^{N}_{a}\rangle$ 
is dynamically generated within a time interval 
of duration $T$ from an initial Hamming-weight-$a$ product state [cf. Eq.~\eqref{init_state}].

\begin{figure}[t!]
\includegraphics[width=0.95\columnwidth]{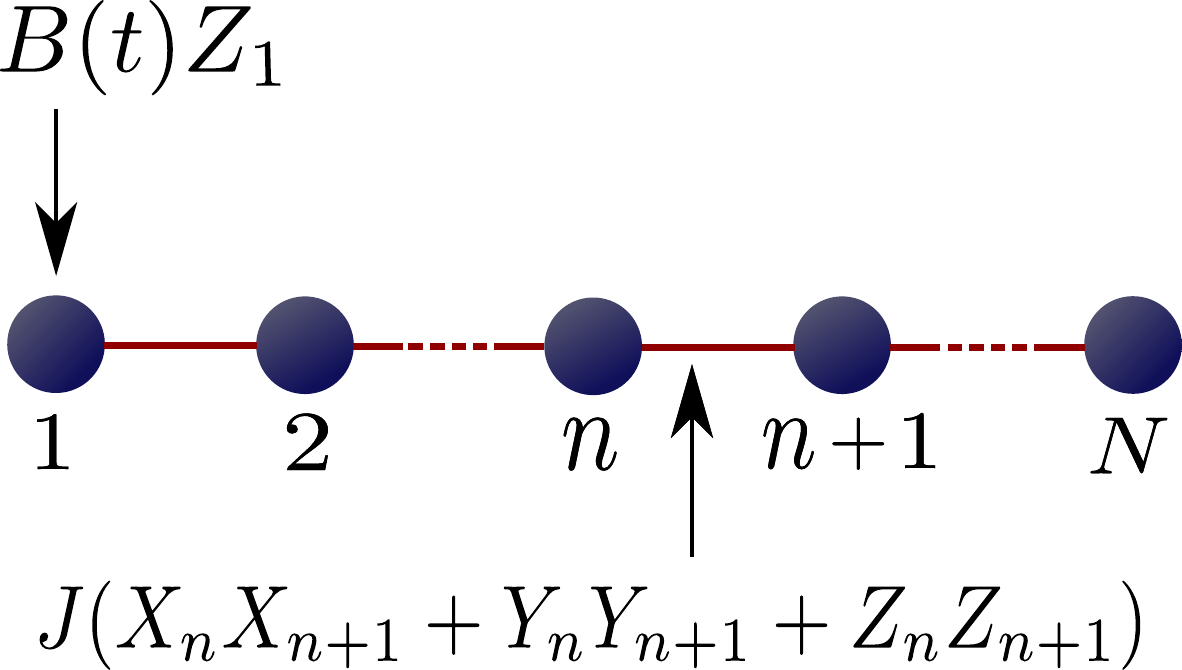}
\caption{\label{CoupledHeisenbergZcontrol:fig4}Schematic illustration of an $N$-qubit array with nearest-neighbor isotropic Heisenberg interaction and a local $Z$ control 
acting on the first qubit (actuator) 
in the array. }
\end{figure}

While, as explained above, the subspace controllability of the system holds for each of the three possible directions of 
$\mathbf{B}(t)$ [cf. Sec.~\ref{SubspControl_XXZandXYZ}],
Dicke states are eigenstates of the $z$ projection $S_z$ of the collective spin operator [cf. Eq.~\eqref{defSz}], 
where the eigenvalue corresponding to the state $|D^{N}_{a}\rangle$ is given by $N/2 - a$. This is what makes the $z$ direction special as far as Dicke states are concerned. Because of that, we will discuss dynamical generation of 
Dicke states with a local $Z$ control on the first qubit 
(for an illustration, see Fig.~\ref{CoupledHeisenbergZcontrol:fig4}), which constitutes 
one special realization of the family of the local-control Hamiltonians in Eq.~\eqref{TotalHamIsoHeisSingleCtrl}. Therefore, the
control part of the total system Hamiltonian  
will hereafter have the form $B(t)Z_1$. The total Hamiltonian of an $N$-qubit array under such circumstances is then given by
\begin{eqnarray}
H(t) = &J&\sum_{n=1}^{N-1}\:\left(X_n
X_{n+1}+Y_{n}Y_{n+1}+Z_{n}Z_{n+1}\right) \nonumber\\
&+& B(t)Z_1 \:. \label{TotalHamIsoHeisZCtrl}
\end{eqnarray}

The dynamical generation of Dicke states in a qubit array described by the Hamiltonian of Eq.~\eqref{TotalHamIsoHeisZCtrl},
with the first qubit playing the role of actuator [i.e., $n_c =1$ in Eq.~\eqref{TotalHamIsoHeisSingleCtrl}],
is investigated in great detail in what follows; we also explore cases in which the actuator is one of the other qubits in an array ($n_c\neq 1$). Our main control objective in the following is to find the shortest possible ($N$ and $a$-dependent) evolution time $T=T_{\textrm{min}}$ that
allows one to realize the desired 
Dicke state $|D^{N}_a
\rangle$, as well as the corresponding 
(state-specific) time dependence of the control field $B(t)$. The control-field magnitude in the problem at hand will be expressed in units of the coupling strength $J$; at the same time, all the relevant times will be expressed in units of $J^{-1}$ (recall that 
$\hbar=1$). 

Needless to say, the appropriate time-dependence $B(t)$ of the control field that allows one to engineer the desired Dicke state $|D^{N}_a\rangle$ within the shortest possible time $T=T_{\textrm{min}}$ -- and this minimal time itself -- do not follow from the above Lie-algebraic results on subspace controllability of Heisenberg-coupled qubit arrays, but ought to be determined by altogether different 
means. To this end, we employ advanced 
methods of quantum optimal control 
(for details, see  Secs.~\ref{MethodsOptCtrl} and \ref{describe_dCRAB} below).

The Dicke-state generation problem under consideration is that of engineering a target state of a closed quantum system -- up to an unimportant global phase -- in a finite time $T$. Therefore, the relevant figure of merit in this problem is the target-state fidelity
\begin{equation}\label{deffidelity}
\mathcal{F}_{t=T}=\big|\langle \psi(t=T)|D^{N}_a\rangle\big|^2\:,
\end{equation} 
i.e., the module squared of the overlap between the actual state $|\psi(t=T)\rangle$ of the system at time $t=T$ and the target Dicke state $|D^{N}_a
\rangle$; the state $|\psi(t=T)\rangle$ is given by $U(t=T)|\psi(t=0)\rangle$, where $U(t)$ is the time-evolution operator of the system, i.e., the one corresponding to the Hamiltonian in 
Eq.~\eqref{TotalHamIsoHeisZCtrl} [for details of the numerical evaluation of $U(t)$, see Sec.~\eqref{MethodsOptCtrl} below]. 

\subsection{Choice of quantum optimal-control scheme}\label{MethodsOptCtrl}
Here, we motivate our choice of
the optimal-control approach to be employed in what follows for finding the time dependence $B(t)$ of the control field that permits the realization of the sought-after Dicke state $|D^{N}_a\rangle$, i.e., the time dependence maximizing the state fidelity [cf. Eq.~\eqref{deffidelity}].

It is a common practice in quantum optimal control to start with piecewise-constant control fields~\cite{Stojanovic+:12}. In that case the time-evolution operator $U(t)$ of the system -- describing its dynamics for $0\leq t\leq T\equiv N_f\Delta t$ -- is given by a product of factors of the form $\exp[-i(H_0+H_C^{(k)})\Delta t]$, which is characteristic of time-independent Hamiltonians; here $H_0$ is the drift Hamiltonian of the system and $H_C^{(k)}$ the (time-independent) control Hamiltonian during the $k$-th time interval of duration $\Delta t$ ($k=1,\ldots,N_f$) in which the control field has a constant value $B_k$. 

Among the quantum optimal-control schemes that rely on piecewise-constant control fields, the most widely used one is based 
on the GRadient Ascent Pulse Engineering (GRAPE) algorithm~\cite{khaneja_optimal_2005}.
The latter iteratively updates the control-field values in each 
piecewise-constant step in order to maximize the relevant figure of merit. Other notable numerical method that employs piecewise-constant control fields is the Krotov algorithm \cite{goerz2019krotov}, distinguished by its enhanced convergence properties. Similar to GRAPE, the Krotov algorithm belongs to the group of gradient-based, open-loop optimization methods.
Finally, it is worthwhile to mention 
the recently proposed geodesic pulse engineering~\cite{lewis_quantum_2025}, which relies on differential programming and geodesics on the Riemannian manifold of SU$(2^N)$.

In addition to gradient-based algorithms for optimal control, there are also approaches that are not gradient-based
and incorporate bandwidth limitation and smooth pulses in a few-parameter Ansatz.
Examples of such approaches are furnished by the CRAB algorithm~\cite{caneva_chopped_2011}, which encodes pulses in a chopped randomized Fourier basis, as well as 
its systematic improvement -- the dCRAB algorithm~\cite{rach_dressing_2015} (see the detailed description in Sec.~\ref{describe_dCRAB} below); another example for optimal-control 
algorithms of this class is gradient 
ascent in function space~\cite{lucarelli_quantum_2018}, which places special emphasis on smooth control fields. Motivated by its superior convergence properties, enhanced compared to its parent CRAB algorithm, in the state-engineering problem at hand we employ the dCRAB algorithm~\cite{rach_dressing_2015}. 

By contrast to the piecewise-constant case, in the case of 
smooth time-dependent control pulses
(continuous-wave approach to quantum optimal control) the total system Hamiltonian $H(t)$ carries an explicit time dependence and, in general, does not commute with itself at different times, i.e., $[H(t'),H(t'')]\neq 0$. Consequently, the time-evolution operator of 
the system is given by the most general form
\begin{equation}\label{eq:tdep-U}
U(t)={\cal T}\exp\left[-i\int_{0}^t H(t') 
{\rm d}t'\right] \:,
\end{equation}
where ${\cal T}$ stands for the time-ordered product of operators.

In what follows, we will be concerned with a quantum state-control problem. Accordingly, instead of computing the time-evolution operator $U(t=T)$ corresponding to a finite-time evolution of the system using a Suzuki-Trotter type decomposition~\cite{Kaldenbach+:24}, we will directly embark on solving the time-dependent 
Schr\"{o}dinger equation (TDSE)
\begin{equation} \label{TDSEideal}
\frac{d}{dt}|\psi(t)\rangle = -i
[H_{XXX}+ H_C(t)]\:|\psi(t)\rangle\:, 
\end{equation}
which governs the dynamical evolution of the 
state $|\psi(t)\rangle$ of our system,
assuming that its initial 
state $|\psi(t = 0)\rangle$ is the 
Hamming-weight-$a$ 
product state in Eq.~\eqref{init_state}. 
We integrate this TDSE up to $t=T$ by making use of an adaptive explicit $5$-th order Runge–Kutta method [embedded $5(4)$ pair]~\cite{Tsitouras:11}. 
Given that this integration method 
does not guarantee that the 
norm of the final state vector is preserved~\cite{NRcBook},
we also carry out the normalization 
step by taking $|\psi(t = T)\rangle = |\psi(t = T)\rangle
/\lVert |\psi_{\textrm{RK}}(t = T)\rangle\rVert$ as the final state of the system, where 
$|\psi_{\textrm{RK}}(t = T)\rangle$ is the output-state vector given by the integrator at $t = T$ and $\lVert \cdot \rVert$ denotes the vector $L_2$-norm.

\subsection{dCRAB: short description of the algorithm}\label{describe_dCRAB}
In what follows, we describe the essential ingredients of the dCRAB algorithm~\cite{rach_dressing_2015}, along with some basic features of our own implementation thereof in the problem under consideration.
These features include, e.g., our approach to 
global optimization -- which is more advanced than the one conventionally used in applications
of the dCRAB formalism -- as well as some quantitative, implementation-related details.

The first issue in any application of the dCRAB formalism is the choice of the functional basis in which to parametrize the control fields. While, in principle, different bases can be employed in conjunction with the dCRAB formalism~\cite{Mueller+:22}, we parametrize our smooth control pulses using a truncated randomized Fourier basis, i.e., a basis that consists of a finite number of sine- and cosine harmonics with randomly sampled frequencies. In this basis, the control field $B(t)$ is decomposed as 
\begin{equation} \label{PulseParameters}
B(t)=\sum_{m=0}^{M-1} \left[\:c_m 
\cos(\omega_m t) + s_m\sin(\omega_m t)\:\right] \:,
\end{equation}
where $\omega_m$ are randomly chosen frequencies within a given interval; here, we take $\omega_m=2\pi(m+r_m)/T$, where $r_m$ is randomly sampled in the interval $[-0.5,0.5]$. As a consequence of the dependence of 
the control field $B(t)$ on the parameters $\{c_{0},c_{1},\ldots, c_{M-1}\}$ and 
$\{s_{0},s_1,\ldots, s_{M-1}\}$, the 
time-evolution operator $U(t)$ of 
the system is also a function of 
those parameters. By extension, the same is true of the state $|\psi(t=T)\rangle$ of the system at $t=T$ and the target-state infidelity $1-\mathcal{F}_{t=T}$ [cf. Eq.~\eqref{deffidelity}].  
While the steps described up to now constitute the standard CRAB optimization, it is worthwhile taking into account that the basis has a finite number of degrees of freedom and the convergence may end in a non-optimal fixed point. The principal idea of dCRAB is then to dress the obtained pulse with further iterations of CRAB, namely by introducing an updated pulse
\begin{equation}
\begin{split}
B^{(l)}(t)=B&^{(l-1)}(t)\ +
\\&+\sum_{m=0}^{M-1} \left[c^{(l)}_m \cos(\omega^{(l)}_mt) + s^{(l)}_m\sin(\omega^{(l)}_mt)\right] \:,
\end{split}
\end{equation}
where $\omega^{(l)}_m$ are newly sampled random frequencies, $B^{(l-1)}(t)$ denotes the pulse obtained at the previous CRAB iteration, and $l$ enumerates repetitions of the 
dressing procedure.
The optimization is performed once again, using $\{c^{(l)}_m,s^{(l)}_m\}$ as optimization parameters  and keeping $B^{(l-1)}(t)$ fixed.
The complete functional form of the pulse in the end reads
\begin{equation}
B(t)=\sum_{l=0}^{L-1}\sum_{m=0}^{M-1}\left[c^{(l)}_m \cos(\omega^{(l)}_mt) + s^{(l)}_m\sin(\omega^{(l)}_mt)\right] \:,
\label{eq:total-pulse-dCRAB}
\end{equation}
where $L$ denotes the total number of dressing iterations. Here, we set a maximum value of $L_{\rm max}=10$, with the possibility of an early termination of the dressing procedure (hence, of the whole optimization as well) if the figure of merit is below a chosen threshold value (here  
$1-{\cal F}_{t=T}=10^{-3}$).
We recall once again that, in the pulse parametrization of Eq.~\eqref{eq:total-pulse-dCRAB}, the optimization parameters $\{c^{(l)}_m,s^{(l)}_m\}$ are not optimized all at once, but for one value of $l$ at a time.

We carry out local searches for the minima of the target-state infidelity $1-{\cal F}_{t=T}$ [cf. Eq.~\eqref{deffidelity}] using the Nelder-Mead simplex method~\cite{NRcBook}. Moreover, to account for the complexity of the optimization landscape in the problem at hand, we also make use of the 
{\em multistart-based clustering} (also known as {\em multistart-based global random search}) algorithm  ~\cite{TornZilinskasBook}. The latter is widely used for finding global minima of objective functions that have a large number of very close local minima~\cite{Stojanovic+Vanevic:08,Stojanovic:20} 
and consists of the following steps.
Firstly, one randomly selects a large 
($\sim 10^3$) sample of points in the candidate-solution space. Secondly, 
one selects a much smaller ($\sim20$) number of points 
that yield the smallest values of the objective function
(here the target-state infidelity) 
and performs local searches for minima around each of them; the one with the smallest value of the objective function is then adopted as the global minimum. The adequacy of this algorithm is corroborated by the stability of the obtained result for the global minimum of the objective function upon altering the number of random points in the initial step of the algorithm. Here, this strategy is used at the beginning of each dressing iteration, to obtain the best initial sample for global minimization.

\begin{figure}[t!]
\centering
\includegraphics[width=8.6cm]{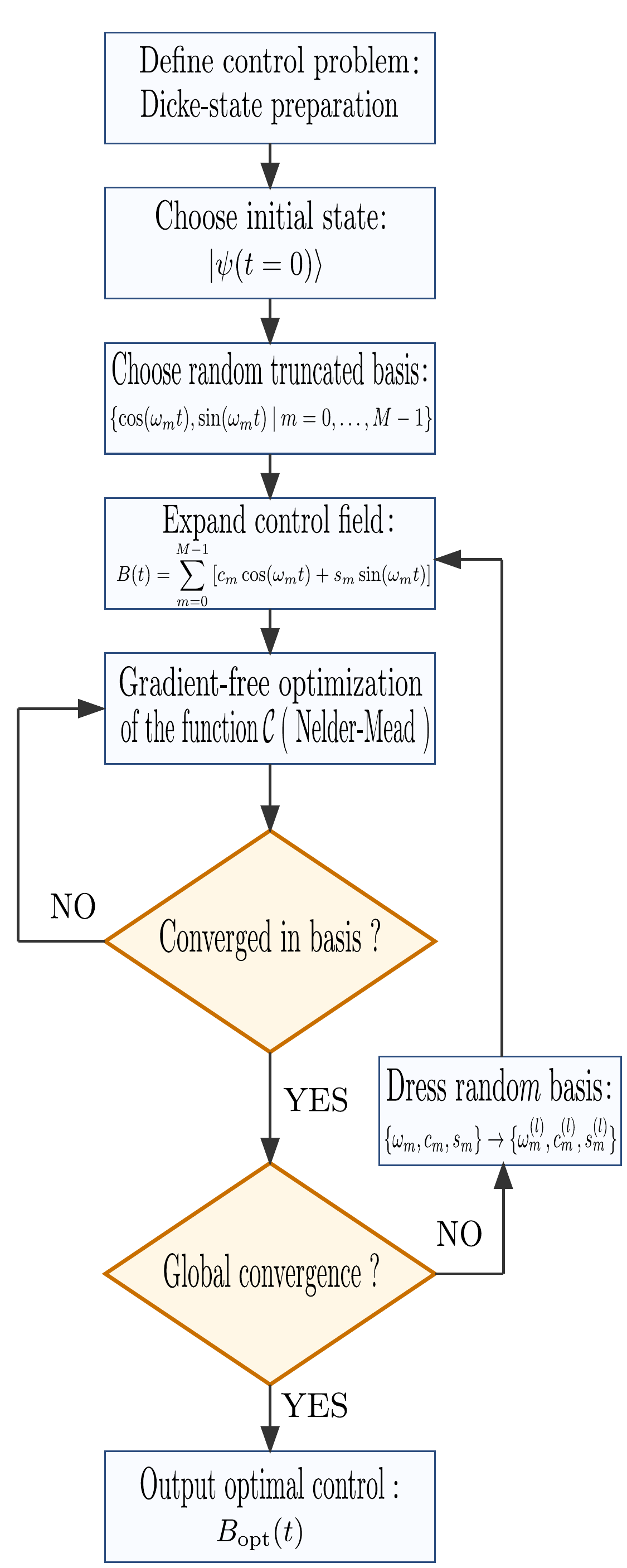}
\caption{Flowchart diagram of the dCRAB algorithm of quantum optimal control, as used in quantum-state engineering. The notation for the cost function and random truncated basis is the one employed in the Dicke-state preparation problem.}
\label{fig:dCRABflow}
\end{figure}

The whole procedure described above allows the encoding of limited-bandwidth pulses and has already proven its effectiveness in realistic experimental settings~\cite{Omran+:19}. We remark that, even if in the problem at hand the maximum bandwidth is tied to the value of $M$ (i.e., increasing $M$ increases the maximum frequency as well), nothing prevents from defining a given frequency interval $[\omega_{\rm min},\omega_{\rm max}]$ and sampling $M$ frequencies within it, hence allowing to freely change the number of degrees of freedom in the parametrization without affecting the targeted interval 
$[\omega_{\rm min},\omega_{\rm max}]$. 

The extent to which 
the use of time-dependent control fields can speed up a certain process depends on the control-field strength.
For definiteness, in the problem at hand we assume that 
the control-field strength $B(t)$ (expressed in units of $J$) during the entire duration of control is within the range $[-4\pi,4\pi]$. To include this constraint in our numerical calculations, as the actual figure of merit we employ the target-state infidelity $1-{\cal F}_{t=T}$ complemented with terms accounting for the penalties incurred when the absolute value of $B(t)/J$ exceeds $4\pi$, i.e.,
\begin{eqnarray} \label{FOM_def}
{\mathcal C} = 1-{\cal F}_{t=T} &+& J^{-1}\big[B_{\rm max}\:\Theta(B_{\rm max}/J-4\pi)\nonumber\\
&-& B_{\rm min}\:\Theta(-4\pi-
B_{\rm min}/J)\big] \:,
\end{eqnarray}
where $\Theta(\ldots)$ denotes 
the Heaviside step function, with
$B_{\rm max}=\max_tB(t)$
and $B_{\rm min}=\min_tB(t)$.

The flowchart diagram of the dCRAB algorithm, illustrating the search for the optimal control field $B_{\textrm{opt}}(t)$ through a gradient-free optimization (here based on the Nelder-Mead simplex method) of the cost function $\mathcal{C}$ over the expansion coefficients of the control field in a dressed random basis, is depicted in Fig.~\ref{fig:dCRABflow}.

\section{Dicke-state preparation
: Results and Discussion} \label{IdealSysResults}
Having explained the essentials of our optimal-control scheme for Dicke-state preparation, 
in what follows we present the results obtained by implementing this scheme.
We first discuss our results obtained for Dicke- and $W$-state generation in Heisenberg-coupled arrays with up to $N=9$ qubits and a local control field acting in the $z$ direction on the first qubit in 
the array, starting from the
initial state in Eq.~\eqref{init_state}   (Sec.~\ref{dCRABbasedDicke}). We then discuss how the obtained results are changed when this last initial state 
is replaced by other Hamming-weight-$a$ product states, as well as for different choices of the actuator qubit (Sec.~\ref{DepInitStateActuator}).

\subsection{dCRAB-based generation of Dicke states} \label{dCRABbasedDicke}
For every target state, we minimized the figure of merit in Eq.~\eqref{FOM_def} over the parameters $\{c^{(l)}_{m},s^{(l)}_{m}\:|\:m=0,\dots,M-1;\:l=0,\dots,L-1\}$, for the varying total duration $T$ of the pulse. In each 
case, the total evolution time $T$ is varied in small steps starting from an initial value -- for which a high target-state fidelity cannot be 
achieved -- until a high-fidelity realization of the desired target state becomes possible; the shortest time $T$ for which 
a realization of the desired state within the assumed fidelity
threshold ($1-\mathcal{F}_{t=T}< 10^{-3}$) is possible is then identified with the shortest
possible state-preparation time $T_{\textrm{min}}$.

\begin{figure}[t!]
\centering
\includegraphics[width=0.95\linewidth]{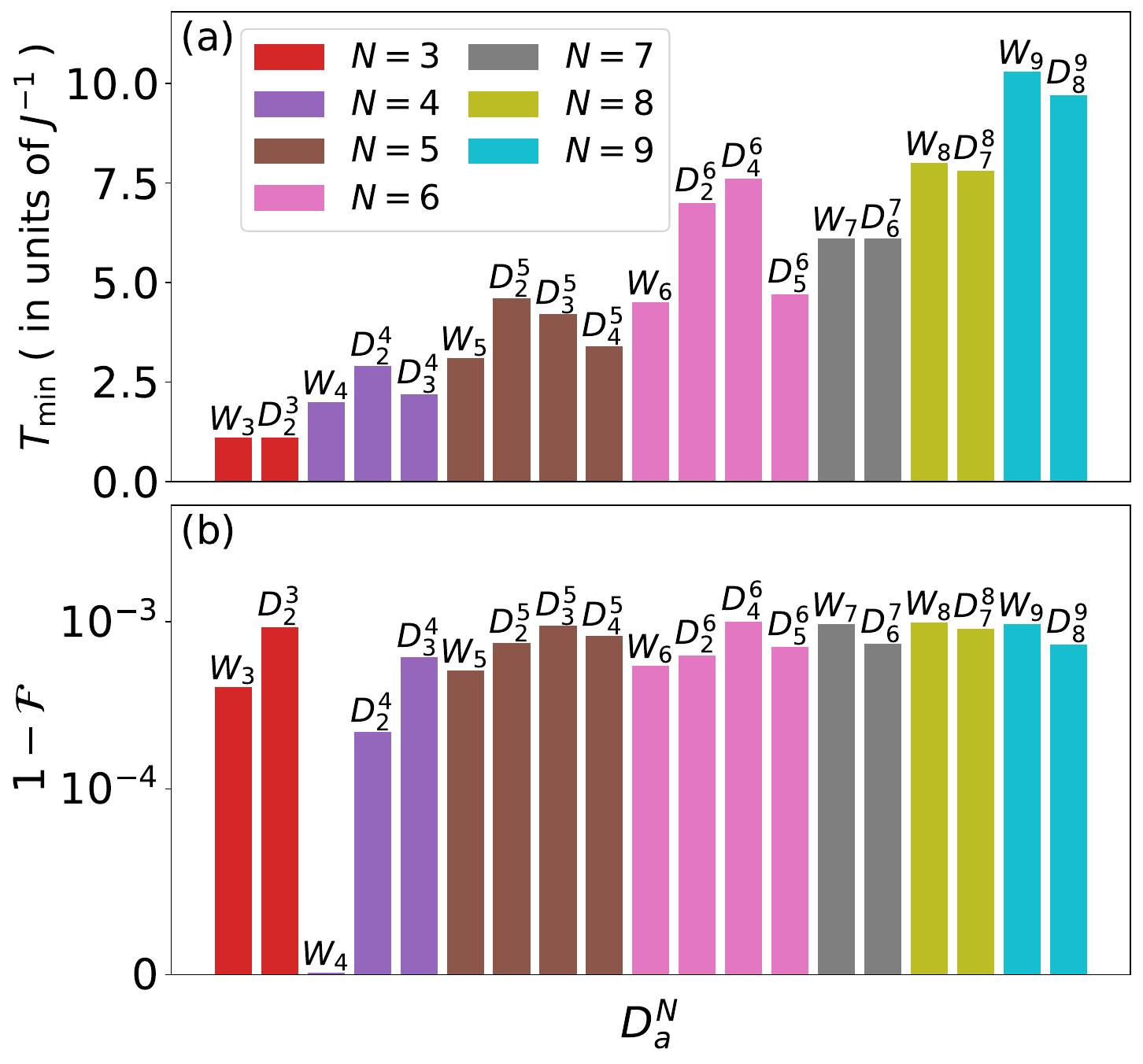}
\caption{Pictorial overview of the 
results obtained using the optimal-control approach based on the dCRAB formalism (with $M=15$ harmonics in the truncated random Fourier basis) for an $N$-qubit array ($N=3,\ldots,9$) with isotropic, 
nearest-neighbor Heisenberg coupling  
and a $Z$ control on the first qubit: 
(a) The shortest possible times $T_{\textrm{min}}$ required for the realization of both 
genuine Dicke states 
($|D^{N}_a\rangle$, with $a\geq 2$) and $W$ states ($|W_N\rangle\equiv|D^{N}_1\rangle$) starting from the initial ($t=0$)
Hamming-weight-$a$
product states in 
Eq.~\eqref{init_state}; 
(b) The highest fidelities (i.e., the smallest infidelities 
$1-\mathcal{F}$) obtained for each of the investigated $N$-qubit states through an evolution of duration $T_{\rm min}$ that starts from the initial states of Eq.~\eqref{init_state}.}
\label{fig:ResultsOverview}
\end{figure}

The results we obtained in the investigated examples of Dicke states, both for the genuine 
Dicke states ($a\geq 2$) and in the special case of $W$ states ($a=1$), are illustrated in 
Fig.~\ref{fig:ResultsOverview}. In particular, Fig.
~\ref{fig:ResultsOverview}(a) shows the obtained times $T_{\textrm{min}}$ required to realize 
particular Dicke states; these times
are in the range between around $J^{-1}$
(for the three-qubit $W$ state) and $10\:J^{-1}$
(for the nine-qubit $W$ state). 
At the same time, Fig. ~\ref{fig:ResultsOverview}(b) depicts
the target-state infidelities achieved for $T=T_{\textrm{min}}$ in the numerical optimization process; these infidelities are mostly between $10^{-3}$ and $10^{-4}$. 

Our optimal-control approach to engineering Dicke states
$|D^{N}_a\rangle$ ($a=1,\ldots,N-1$) in an $N$-qubit array entails working 
in a computational basis that consists of $\uber{N}{a}$ states -- the dimension of the $a$-excitation subspace $\mathcal{H}_a$ of the total $2^{N}$-dimensional Hilbert space (recall 
Sec.~\ref{SubspControl_XXZandXYZ} above). Given that for 
the fixed system size $N$ the dimension of this subspace 
grows rapidly with increasing the excitation 
number $a$ -- before reaching the maximal value 
$N!/(\lfloor N/2 \rfloor !)^2$ for $a=\lfloor N/2 \rfloor$ [which is equal to
$N/2$ for even values of $N$, and $(N-1)/2$ for odd ones] -- and, 
for large $N$, even becomes asymptotically 
exponential in $N$ (cf. Sec.~\ref{SubspControl_XXZandXYZ}), the underlying numerical-optimization problem [namely, finding the global minimum of the generalized figure of merit in Eq.~\eqref{FOM_def}] becomes increasingly more demanding from the computational standpoint. As a consequence, the numerical burden involved in engineering Dicke states 
within our adopted optimal-control scheme becomes computationally prohibitive for large system sizes $N\gtrsim 10$ and values of $a$ close to $N/2$.

The level of difficulty of optimal control-based realizations of different multiqubit states in a qubit array that is subject to external time-dependent control fields is not necessarily related only to the entanglement content of those states, even if the initial state is usually assumed to be a product state. Loosely speaking, the complexity of the optimal control-based realization of a certain target multiqubit state depends on its distance from naturally generated dynamical manifolds; on a more formal level, this complexity is largely determined by the overlap between the DLA of a quantum system under consideration -- which, in turn,  depends on
the type of interaction between qubits in the system and the form of its coupling with the external control field (cf. Sec.~\ref{basics}) -- and the 
target-state stabilizer structure. In particular, if the Hamiltonian that governs the system dynamics naturally generates trajectories toward the target state, the accompanying optimization process is usually straightforward.
For instance, collective (all-to-all) Ising-type qubit-qubit interactions naturally generate Greenberger-Horne-Zeilinger (GHZ) type states $|\textrm{GHZ}_N\rangle = (|0\rangle^{\otimes N}+|1\rangle^{\otimes N})/\sqrt{2}$, which are highly entangled yet structurally fairly 
simple as they belong to the subspace span$\{|00\ldots 0\rangle, |11\ldots 1\rangle\}$, i.e., their support in the $2^N$-dimensional Hilbert space $(\mathbbm{C}^2)^{\otimes N}$ is equal to $2$. In Ref.~\cite{Omran+:19}, such states were realized by employing a dCRAB-based optimal-control scheme in systems with up to $20$ qubits whose logical $|0\rangle$ and $|1\rangle$ states are encoded in the neutral-atom ground state and a highly-excited Rydberg state, respectively; qubits in such a system have Ising-type interactions. For $W$ states, the $XX$ (flip-flop) 
qubit-qubit interaction plays an analogous role to the 
Ising interaction in the case of GHZ states. By contrast, here we are faced with the task of generating both $W$ and genuine Dicke states in qubit arrays with the isotropic Heisenberg ($XXX$) interaction, where the dimension of their 
two-excitation invariant subspace for an $N$-qubit array already scales as $N^2$, hence the limitation to the relatively small system size ($N\leq 9$).

\begin{figure}[t!]
\centering
\includegraphics[width=0.95\linewidth]{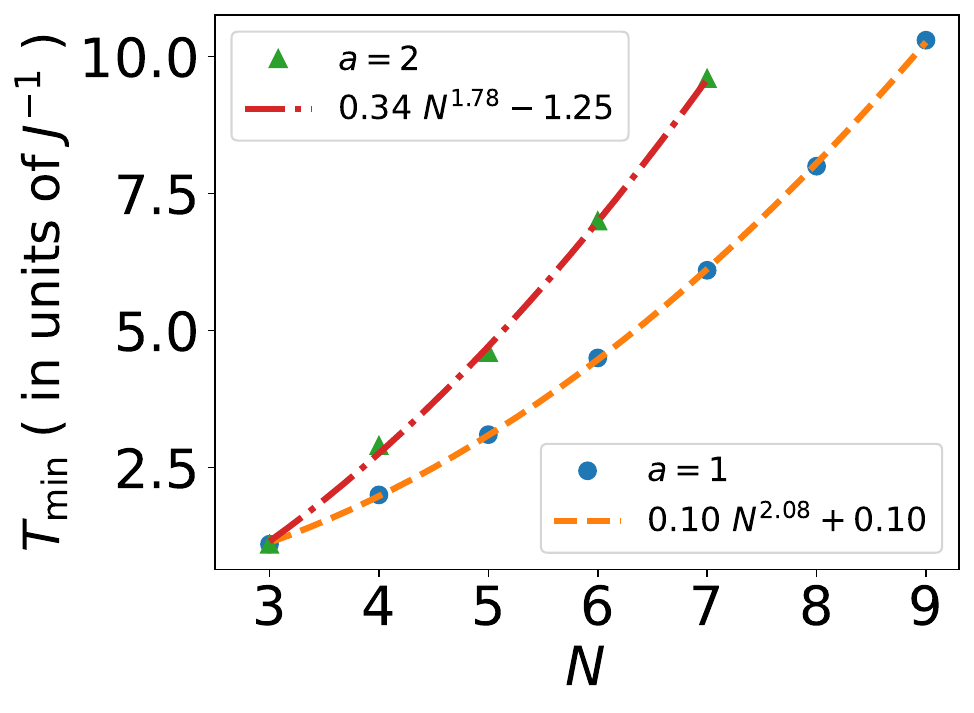}
\caption{Illustration of the scaling of 
the shortest possible state-preparation 
times $T_{\textrm{min}}$, corresponding to 
the initial ($t=0$) 
Hamming-weight-$a$
product states in 
Eq.~\eqref{init_state}, 
for both $W$ states 
$|W_N\rangle\equiv|D^{N}_{a=1}\rangle$
and two-excitation Dicke states $|D^{N}_{a=2}
\rangle$, with the number $N$ of qubits. The two scaling laws $T_{\textrm{min}}(N)$ are established by fitting the results obtained using the dCRAB formalism, with $M=15$ harmonics in the truncated random Fourier basis.}
\label{fig:ScalingOfTmin}
\end{figure}

In order to quantitatively assess the potential practical utility of the proposed local-control approach in engineering highly-entangled multiqubit states of Heisenberg-coupled qubit arrays it is of interest to deduce the scaling of the shortest possible state-preparation times $T_{\textrm{min}}$ for both $W$ states and genuine Dicke states with the number of qubits
$N$. Based on our obtained numerical results, a fitting procedure aimed at extracting a power-law dependence of $T_{\textrm{min}}$ on $N$ yields the result
\begin{equation} \label{scaleTmin_aeq1}
T_{\textrm{min}}(N)/J^{-1}= 0.10\times N^{2.08}+0.10
\end{equation}
for $N$-qubit $W$ states ($a=1$), and
\begin{equation}\label{scaleTmin_aeq2}
T_{\textrm{min}}(N)/J^{-1}= 0.34\times N^{1.78}-1.25 
\end{equation}
for Dicke states $|D^{N}_{a=2}\rangle$. Therefore, the scaling of the shortest possible state-preparation time $T_{\textrm{min}}$ obtained using our optimal-control approach based on the dCRAB algorithm with $N$ is approximately quadratic in $N$ for $W$ states and slower than quadratic for $a=2$ Dicke states (for an illustration, see Fig.~\ref{fig:ScalingOfTmin}). The two curves in Fig.~\ref{fig:ScalingOfTmin} have an intersection for $N=3$ because the states 
$|D^3_1\rangle\equiv|W_3\rangle$ and $|D^3_2\rangle$ are equivalent up to the interchange $|0\rangle \rightleftarrows|1\rangle$ of the single-qubit 
basis states (recall Sec.~\ref{DickeStatesReview}), 
hence the equal preparation times obtained. 

It is pertinent at this point to comment 
on the obtained dependence of $T_{\textrm{min}}$ on $N$ for Dicke states. It has become well-known by now that an overwhelming majority of the as yet proposed analog schemes for the
Dicke-state preparation in various physical
systems are characterized by state-preparation times that exhibit superlinear dependence on
$N$~\cite{Keating+:16}. In 
this context, our principal 
finding -- the 
obtained $\mathcal{O}(N^{1.78})$ scaling 
of $T_{\textrm{min}}(N)$ with 
$N$ --  represents a rather favorable scaling for an analog Dicke-state preparation scheme. This is especially true given that this last result 
is obtained by means of what constitutes -- from the controllability standpoint -- the minimal possible 
control resource that allows the realization of the sought-after states in an entire class of systems (interacting qubit arrays) -- a local $Z$ control acting on a single actuator qubit. Needless to say, more conventional scenarios of quantum control 
in interacting qubit arrays in state-of-the-art QC platforms either entail the use of global control fields~\cite{Aiudi+:26} or time-dependent local control fields acting on each qubit in the array~\cite{Arenz+Rabitz:18}.

While Heisenberg-coupled qubit arrays can be realized in other types of physical platforms (e.g., with neutral atoms),
the most straightforward physical realization
of systems of the kind discussed here is the one based on semiconductor spin qubits. 
In this context, it is pertinent to mention that the largest hitherto reported fully-operational semiconductor 
spin-qubit arrays involve only $6$-$8$ qubits; due to control wiring complexity, crosstalk and noise,
as well as readout challenges, these systems are difficult to scale up. For instance, a tunable, coherently controlled $8$-dot linear array of silicon spin qubits has quite recently been demonstrated~\cite{Nickl+:25}; it is worthwhile to mention that
a somewhat larger, $12$-qubit system with nearest-neighbor exchange coupling has also been recently reported~\cite{George+:25} -- however,
in that system two-qubit gates have been demonstrated only between some, but not all, pairs of qubits and maintaining coherent superpositions across all $12$ qubits proved to be nontrivial. Therefore, the system size for which the Dicke-state preparation is discussed in the present work ($N\leq 9$) essentially coincides with that of the largest fully operational spin-qubit arrays reported to date. While the always-on exchange-coupling strength $J$ in such arrays are typically in the range $J/2\pi\sim 10$\:MHz $-$ $1$\:GHz, the ideal scenario for our present purposes is the one with $J/2\pi\sim 50$\:MHz (note that the attendant characteristic timescale is then $J^{-1}\approx 3.2$ ns), with the corresponding control-field magnitude $B(t)/2\pi\sim 0 - 600$\:MHz [recall the assumption pertaining to the range of values of the ratio $B(t)/J$ that was stipulated in Sec.~\ref{describe_dCRAB}]. Rapidly varying control fields of this type can be generated using a micromagnet and an arbitrary waveform generator (AWG). When used together, they allow one to convert a fast electrical pulse (generated by the AWG) into an effective local magnetic control field on a selected spin qubit (the actuator qubit in our envisioned setup).

Given that our obtained numerical results for the shortest possible Dicke-state preparation times (cf. Fig.~\ref{fig:ResultsOverview}) correspond to Heisenberg-coupled qubit arrays with up to $9$ qubits, it is appropriate to comment at this point on 
the anticipated behavior of $T_{\textrm{min}}(N)$ for the two classes of states displayed in Fig.~\ref{fig:ScalingOfTmin} in the case of larger qubit arrays ($N\geq 10$); this is of interest in the context of 
NISQ-level QC architectures, which typically feature between $50$ and a few hundred qubits~\cite{PreskillNISQ:18}. By extrapolating the values of $T_{\textrm{min}}$ using 
Eqs.~\eqref{scaleTmin_aeq1} and \eqref{scaleTmin_aeq2} it is straightforward to 
find that for $N=50$ the shortest times required for realizing $W$ states and $a=2$ Dicke states are $343.5\:J^{-1}$ and $358.2\:J^{-1}$, respectively. For the aforementioned typical value of $J$, these last times are of the order of a few $\mu$s, while typical coherence times 
$T_2$ of spin qubits that can be utilized for the realization of systems of the type considered here are in the range between $100\:\mu$s and a few ms. Therefore, our control-based engineering of Dicke states in a Heisenberg-coupled qubit array should be feasible even in anticipated NISQ-level spin-qubit arrays once fully operational systems of this type are experimentally demonstrated.

Generally speaking, the question of 
obtaining the minimal evolution time 
of a quantum system subject to external control fields that makes a desired state transformation of the system possible is inextricably linked to the concept of  
{\em quantum speed limit} ~\cite{Deffner+Campbell:17}. 
While the latter concept was originally 
thought of as that of an intrinsic timescale of unitary quantum dynamics, 
subsequent studies explicitly showed the connection between 
this concept and optimal control~\cite{Caneva+:09}. To be more specific, the quantum speed limit 
was shown to yield the shortest timescale required for quantum 
optimal-control algorithms to converge; this finding was a 
{\em de-facto} demonstration that quantum speed limits are attainable and provided 
a definition of optimality in the quantum-control context.
In the present work, we obtain the shortest times 
for the generation of highly-entangled Dicke states from an initial product state in accordance with this same criterion of optimality applied in conjunction with the dCRAB algorithm
of quantum optimal control. Therefore, the present state-engineering study can be seen as that of generating Dicke states in Heisenberg-coupled qubit arrays with a single local control at the quantum speed limit.

\begin{figure}[b!]
\centering
\includegraphics[width=0.95\linewidth]{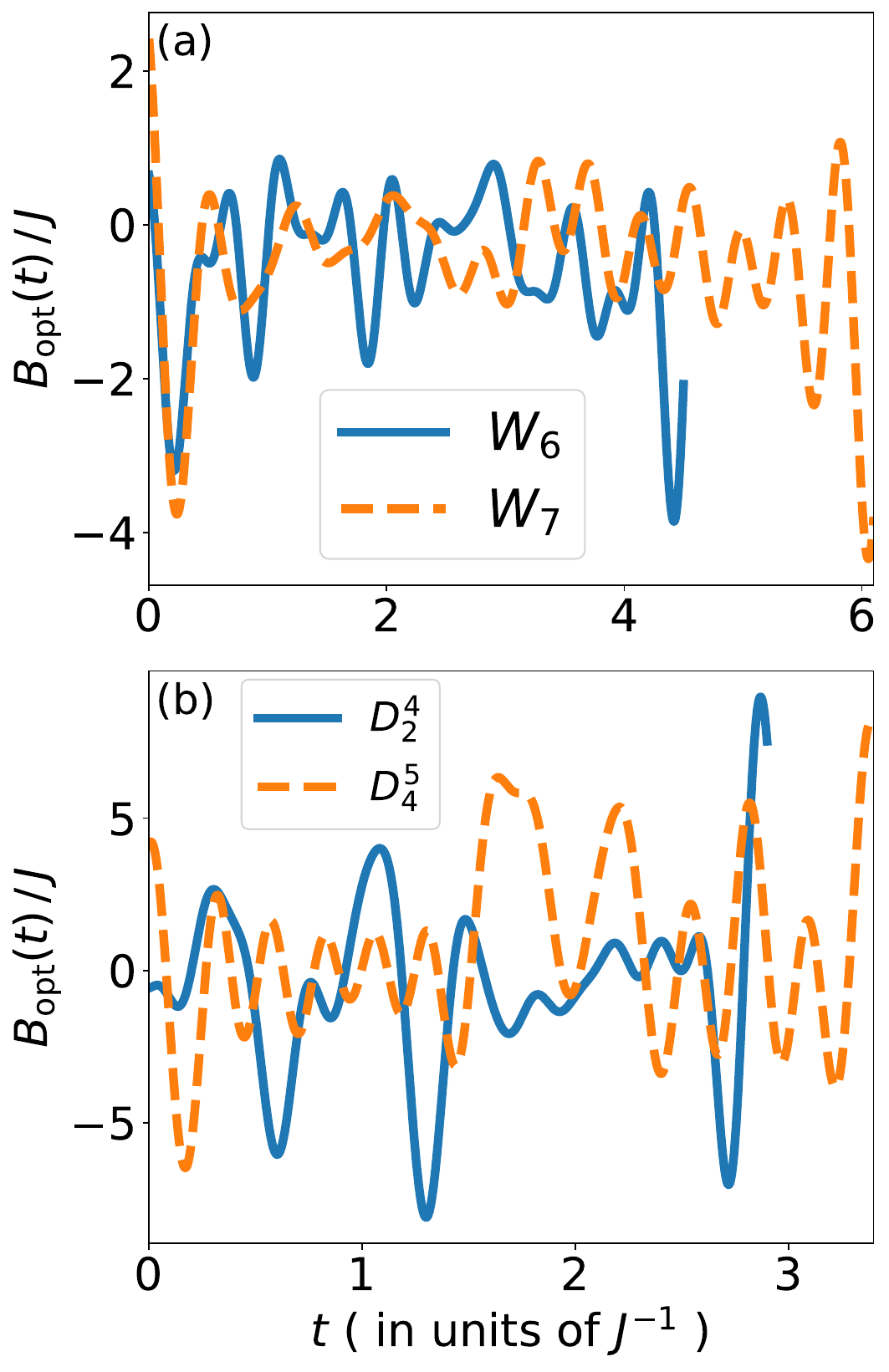}
\caption{Optimal control fields $B_{\textrm{opt}}(t)$ that enable the realization of (a) $W$ states 
$|W_6\rangle$ and $|W_7\rangle$, and 
(b) Dicke states  $|D^4_2\rangle$ and 
$|D^5_4\rangle$, starting from the initial ($t=0$) product states in 
Eq.~\eqref{init_state}. These 
results are obtained using the dCRAB formalism, with $M=15$ harmonics in the truncated random Fourier basis.}
\label{fig:TimeDepCtrl}
\end{figure}

The characteristic time dependencies of optimal control fields 
$B_{\textrm{opt}}(t)$ 
that enable the realization 
of Dicke and $W$ states in the system under consideration are illustrated in Fig.~\ref{fig:TimeDepCtrl}. The examples shown correspond to the 
$W$ states $|W_6\rangle$,$|W_7\rangle$ and 
Dicke states $|D^4_2\rangle$,$|D^5_4\rangle$; their corresponding state-preparation times are all in the range between $2.9\:J^{-1}$ and $6.1\:J^{-1}$. What is evident from Fig.~\ref{fig:TimeDepCtrl} is that the obtained optimal control fields are varying rapidly with time. It can also be inferred from this plot is that the control-field strengths (expressed in units of $J$) $B_{\rm opt}(t)/J$ at all times $0\leq t\leq T_{\textrm{min}}$ do not closely approach the 
upper- and lower bounds of the adopted range $[-4\pi,4\pi]$ of values; this underscores the importance of including the penalty terms in the generalized figure of merit ${\mathcal C}$ [cf. Eq.~\eqref{FOM_def}] that we utilize in 
this problem and {\em a posteriori}
corroborates the fact that
we adopted a sufficiently large range of values of $B(t)/J$.

\begin{figure}[b!]
\centering
\includegraphics[width=0.95\linewidth]{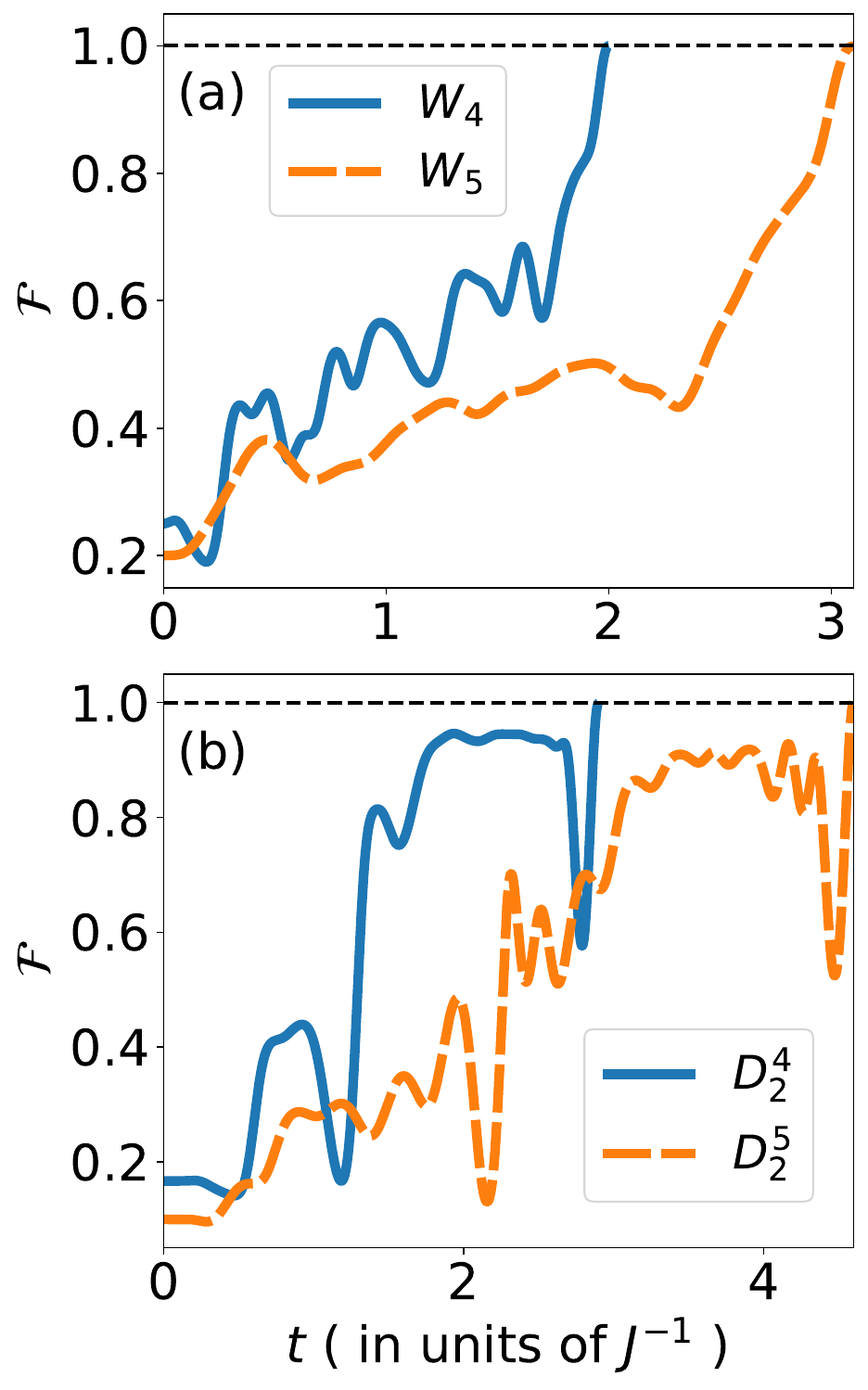}
\caption{Time dependence of the target-state fidelity 
$\mathcal{F}(t)$ throughout the dynamical evolutions 
that start ($t=0$) from the product 
states of Eq.~\eqref{init_state} and end with Dicke- or $W$ states
at $t=T_{\textrm{min}}$. The 
examples shown here correspond 
to (a) $W$ states $|W_4\rangle$ and $|W_5\rangle$, and (b) 
two-excitation Dicke states 
$|D^4_2\rangle$ and $|D^5_2\rangle$.}
\label{fig:fid-evolution}
\end{figure}

Regarding the practical feasibility of implementing rapidly varying control fields, such as those shown in Fig.~\ref{fig:TimeDepCtrl}, a comment is in order here. In present-day solid-state QC systems (based on superconducting- or spin qubits) operating in the microwave regime~\cite{Hansen+:21,George+:25}, shaped control pulses are typically obtained using AWGs); such
devices are presently available with 
the clock jitter of $50$ ps and 
sub-nanosecond time resolutions. 
While in the conventional approach to control-pulse synthesis AWGs only generate a baseband signal and 
a sought-after pulse is obtained through an upconversion to microwave frequencies by mixing with a carrier, the very high currently achievable sampling rates of 
AWGs [exceeding $100$ gigasamples per second (GSa/s)] combined with analog bandwidths in the tens of GHz (indicating the frequency content that an AWG can reproduce accurately, i.e., without distortions) already 
allow direct digital synthesis of microwave pulses. This obviates the need for additional microwave generators even in situations where
precise waveform shaping is required, i.e., for generating complex custom waveforms.

\begin{figure}[b!]
\centering
\includegraphics[width=0.95\linewidth]
{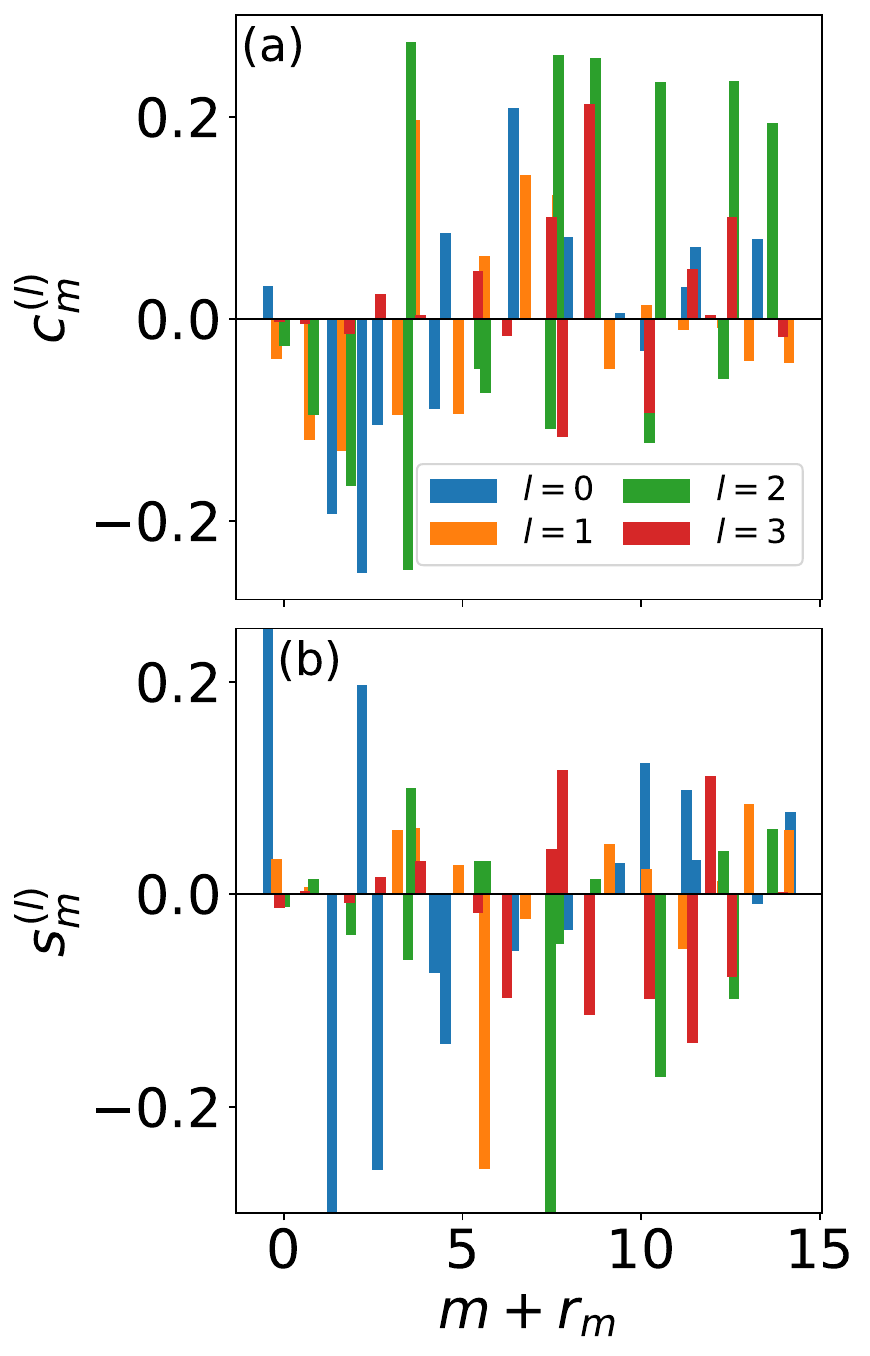}
\caption{Optimal parameters (a) $c^{(l)}_m$, 
and (b) $s^{(l)}_m$,
obtained using the dCRAB formalism. 
The displayed results correspond to the state $|W_6\rangle$, realized starting from the product 
state of Eq.~\eqref{init_state}
at $t=0$, and are obtained with $M=15$ harmonics in the truncated random Fourier basis. Each color is associated with one specific value of $l$, i.e., one specific dressing iteration.}
\label{fig:w5-components}
\end{figure}

Fig.~\ref{fig:fid-evolution} illustrates the time
dependence of the state fidelity $\mathcal{F}(t)$ for both $W$- and genuine Dicke states. For the $a$-excitation Dicke state $|D^{N}_a\rangle$, the dynamical evolution starts 
from the Hamming-weight-$a$ product state in Eq.~\eqref{init_state}; therefore, the initial value of the target-state fidelity is
$\mathcal{F}_{t=0}=\uber{N}{a}^{-1}$. The fidelity has a rather complex time dependence -- this being a consequence of the rapidly-varying control 
field $B(t)$ (cf. Fig.~\ref{fig:TimeDepCtrl}) -- before reaching a value very close to unity at $t=T_{\textrm{min}}$.

\begin{figure}[t!]
\centering
\includegraphics[width=0.95\linewidth]
{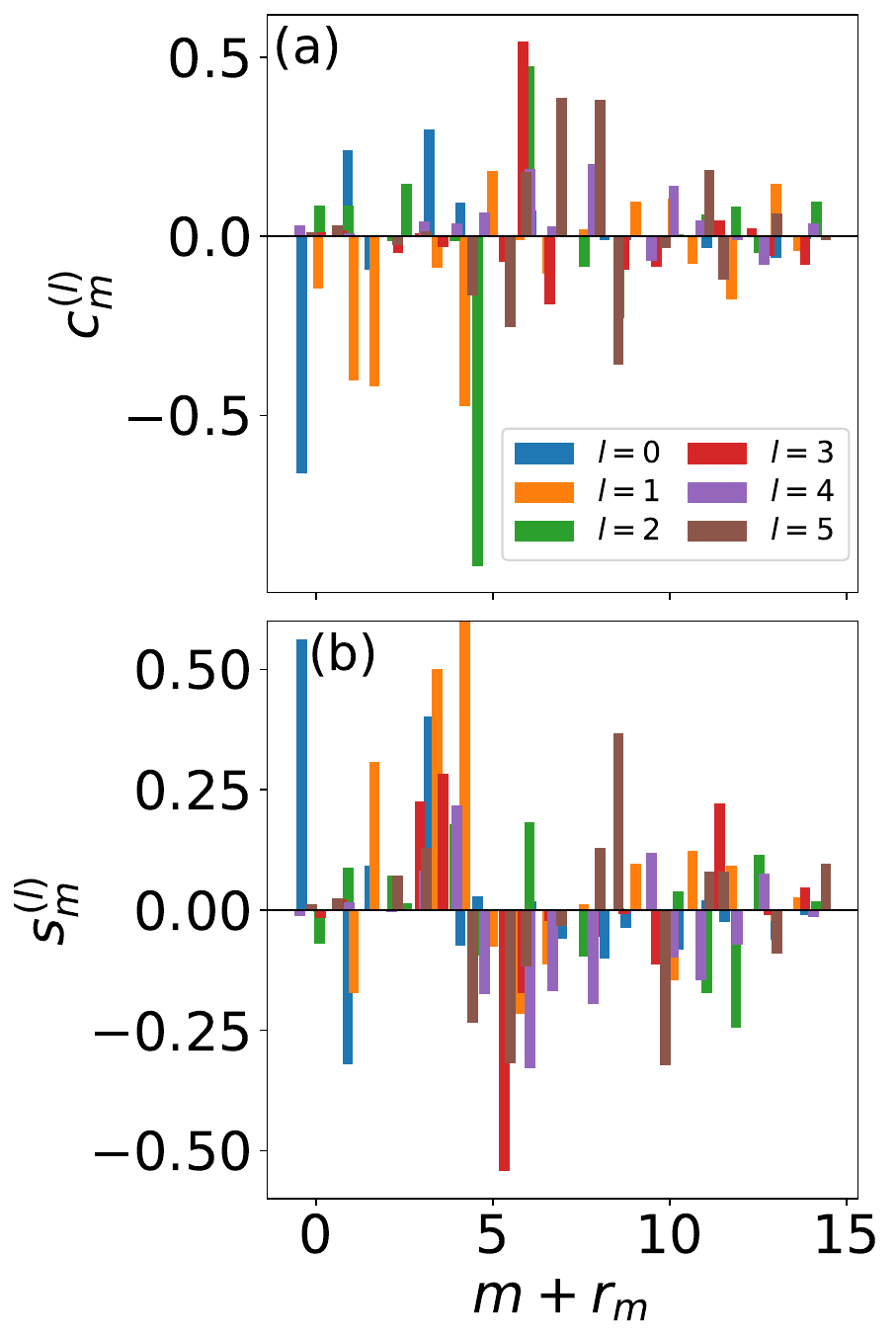}
\caption{Optimal parameters (a) $c^{(l)}_m$, 
and (b) $s^{(l)}_m$, obtained 
using the dCRAB formalism. The
displayed results correspond to the state $|D^4_2\rangle$, realized starting from the product 
state of Eq.~\eqref{init_state}
at $t=0$, and are obtained with $M=15$ harmonics in the truncated random Fourier basis. Each color is associated with one specific value of $l$, i.e., one specific dressing iteration.}
\label{fig:d42-components}
\end{figure}

In our numerical optimal-control calculations we utilize
$M=15$ harmonics in the truncated 
Fourier expansion of Eq.~\eqref{PulseParameters}. 
Regarding the dependence of the obtained results on $M$, the following remark can be made. It 
transpires from our numerical calculations that the obtained results depend 
on $M$ only in an implicit fashion. To be more specific, a sufficiently large value of 
$M$ -- i.e., a large enough number of harmonics 
in the expansion of the control field 
[cf. Eq.~\eqref{PulseParameters}] -- is 
necessary for high-fidelity realizations 
of Dicke states; in particular, the states 
$|D^{N}_a\rangle$ that correspond to 
larger values of $N$ and $a$ cannot be successfully engineered if $M$ is too 
small. However, provided that $M$ is chosen 
to be sufficiently large (as is the case for $M=15$ in the problem under consideration), the actual minimal state-engineering times for different Dicke states do not show any dependence on $M$.

In order to provide a further quantitative characterization of our implementation of the dCRAB formalism, it is instructive to look more closely into the parameters $c^{(l)}_m$ and $s^{(l)}_m$ ($l=0,\ldots,L-1$) resulting from the dressing iterations (cf. Sec.~\ref{describe_dCRAB}).
In Figs.~\ref{fig:w5-components} and \ref{fig:d42-components} these
parameters are displayed for the states $|W_6\rangle$ and 
$|D^4_2\rangle$, respectively. What can be inferred 
from these plots is that all dressing 
iterations [i.e., their corresponding optimization parameters $c^{(l)}_m$ and $s^{(l)}_m$ for $l=1,\ldots,L-1$] play an important role in the optimization process, not only the original optimization parameters 
$c^{(0)}_m \equiv c_m$ and $s^{(0)}_m \equiv s_m$ [i.e., coefficients in the original truncated random Fourier expansion of Eq.~\eqref{eq:total-pulse-dCRAB}].
This important conclusion constitutes an {\em a posteriori} justification for our use of the dCRAB formalism, i.e.,
our preference for this method rather
than its parent CRAB method.

\subsection{Dicke-state preparation for different initial product states and actuator qubits} \label{DepInitStateActuator}

In Sec.~\ref{dCRABbasedDicke} above, we discussed the optimal control-based preparation of Dicke states in Heisenberg-coupled qubit arrays starting from one specific initial product state [cf. Eq.~\eqref{init_state}] and one concrete choice of the actuator qubit ($n_c = 1$).

For the sake of completeness, here we discuss how the obtained results for the shortest possible state-preparation time 
$T_{\textrm{min}}$ depend on the choice of the initial product state and actuator qubit, using for this purpose the Dicke states 
$|D^{4}_2\rangle$ and $|D^{5}_3\rangle$ (for an illustration, 
see Fig.~\ref{fig:act-perms} below). To this end, we repeat our dCRAB-based 
optimal-control scheme for all $6$ possible initial product states in the case of the 
state $|D^{4}_2\rangle$ and all
$4$ choices 
of the actuator qubit [Fig.~\ref{fig:act-perms}(a)]; likewise, we apply this scheme to all $10$ possible initial product 
states and $5$ choices of the actuator in the 
case of the state $|D^{5}_3\rangle$ [Fig.~\ref{fig:act-perms}(b)].

\begin{figure}[b!]
\centering
\includegraphics[width=0.95\linewidth]
{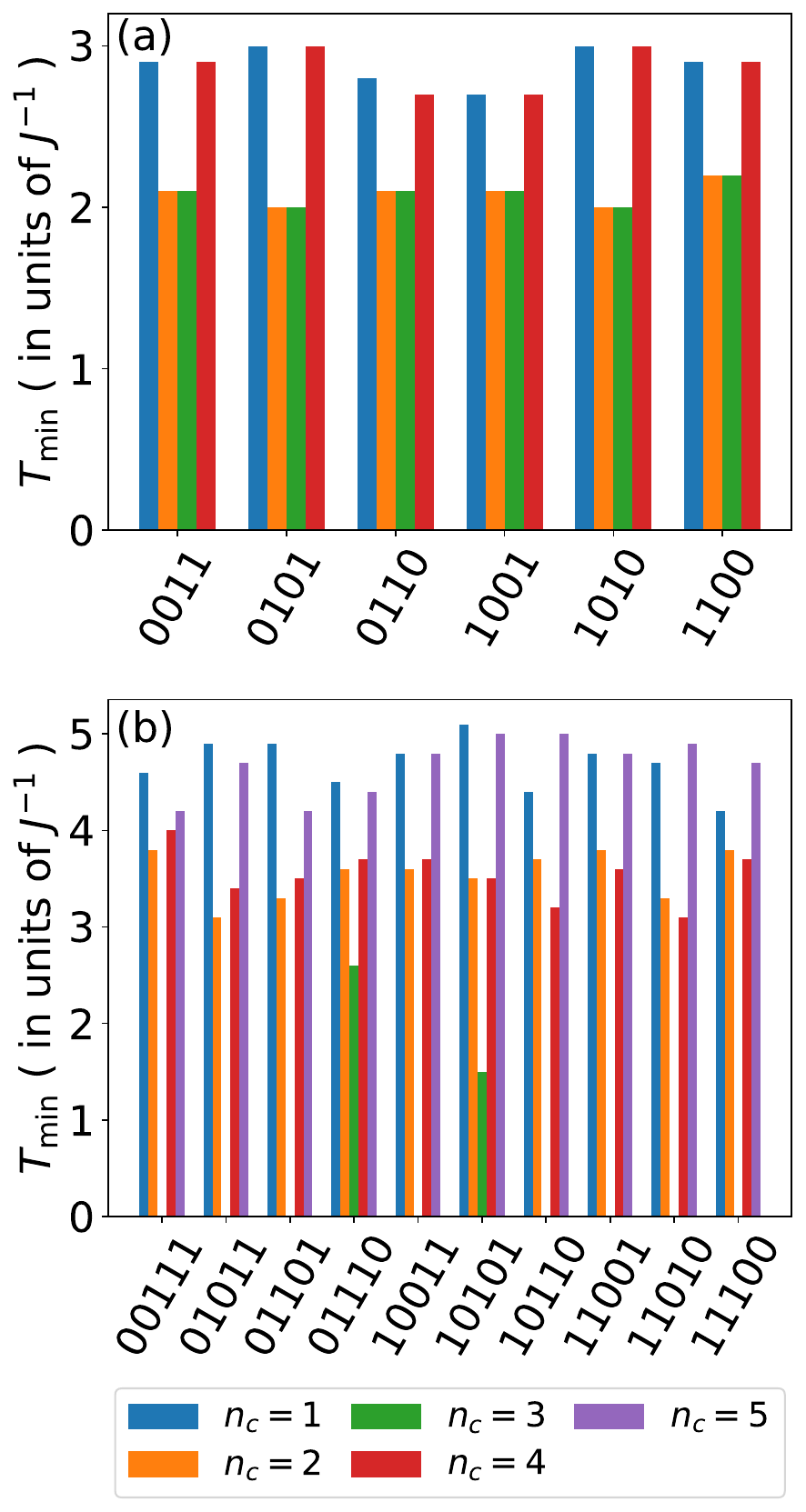}
\caption{Pictorial overview of the numerical results obtained for the shortest possible times $T_{\textrm{min}}$ required to engineer the states (a) $\lvert D^4_2\rangle$, and (b) $\lvert D^5_3\rangle$, of Heisenberg-coupled qubit arrays, using the above dCRAB-based optimal-control scheme with $M=15$ harmonics in the truncated random Fourier basis. The results illustrated here correspond to different choices of the initial product state of a fixed Hamming weight (with all possible such states being listed along the $x$ axes of the two bar graphs) and the actuator qubit $n_c$ (with different colors of the bars corresponding to different values of $n_c$).}
\label{fig:act-perms}
\end{figure}

\begin{figure}
\centering
\includegraphics[width=0.95\linewidth]{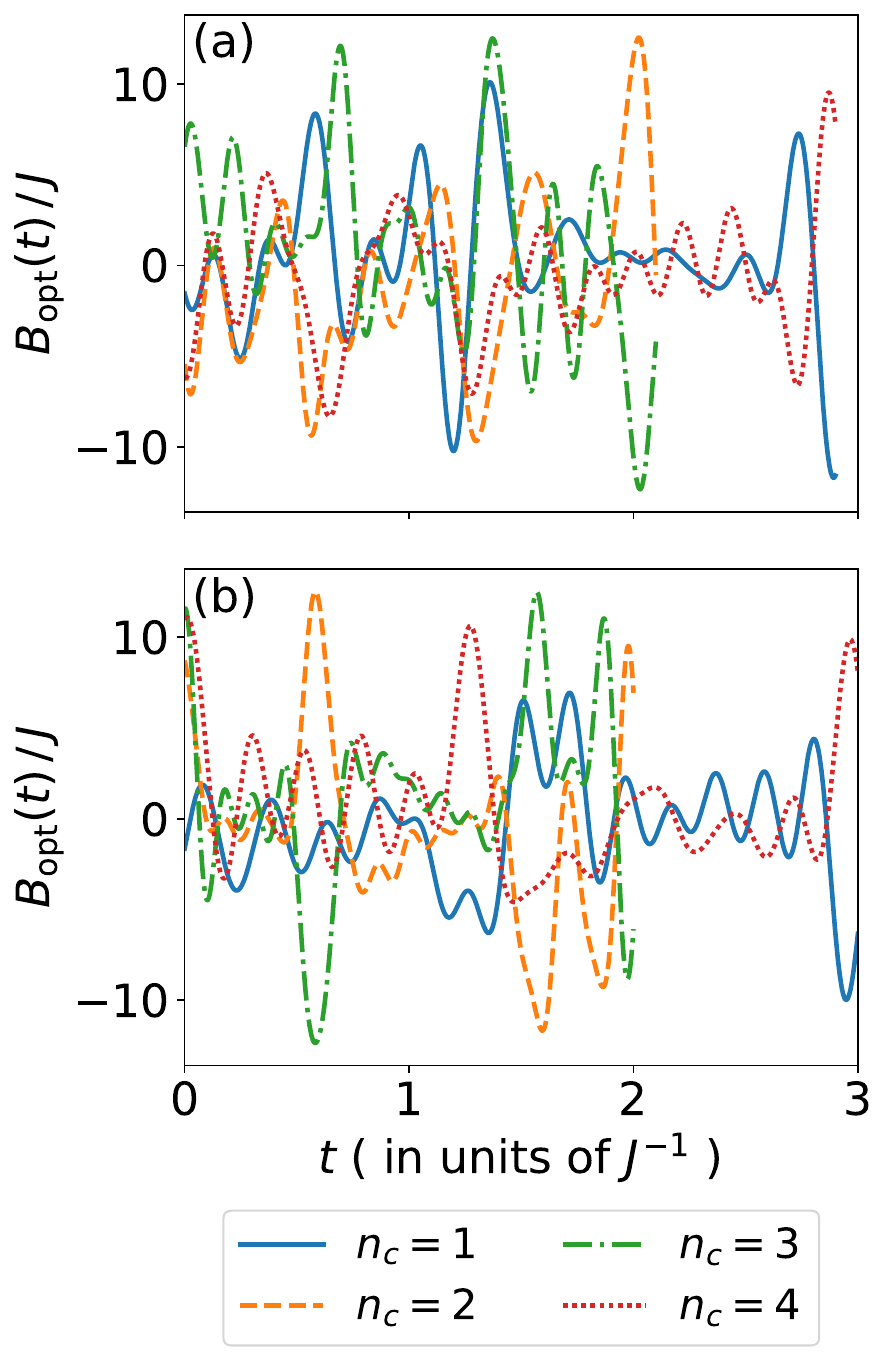}
\caption{Optimal control fields 
$B_{\textrm{opt}}(t)$, expressed in units of $J$, that correspond to the dCRAB-based generation (with $M=15$ harmonics in the truncated random Fourier basis) of the state $|D^{4}_2\rangle$ starting from the Hamming-weight-$2$ product states (a) $|0011\rangle$, and (b) 
$|1010\rangle$. The results displayed here correspond to all possible choices of the actuator qubit $n_c$.}
\label{fig:act-perms-pulses}
\end{figure}

What can be inferred from Fig.~\ref{fig:act-perms} is that $T_{\textrm{min}}$ depends strongly on the position of the actuator within a qubit array -- i.e., the obtained values of $T_{\textrm{min}}$ differ significantly for different choices of
$n_c$; for instance, in the case of engineering the state $|D^{4}_{2}\rangle$ [cf. Fig.~\ref{fig:act-perms}(a)] starting from 
the initial product states $|0101\rangle$ and $|1010\rangle$ we find that $T_{\textrm{min}}$ is around 
$2\:J^{-1}$ for $n_c=2$ and $n_c=3$, while for $n_c=1$ and $n_c=4$ it reaches values of around $3\:J^{-1}$. At the same time, for each fixed position of the actuator (i.e., fixed $n_c$) $T_{\textrm{min}}$ shows a somewhat weaker dependence on the initial product state of fixed Hamming weight. While the lack of permutational symmetry of the total Hamiltonian of the system explains why the obtained times $T_{\textrm{min}}$ for different initial product states are not all equal, the fact that they are still fairly similar can be made plausible based on the following circumstances. Firstly, each of these initial Hamming-weight-$a$ product states $|\psi(t=0)\rangle$ has exactly the same overlap $\langle\psi(t=0)|D^{N}_{a}\rangle={N\choose{a}}^{-1/2}$  with the sought-after Dicke state (hence having the 
same Fubini-Study distance~\cite{Deffner+Campbell:17} from it). Secondly, assuming that the position of the actuator ($n_c$) is fixed, for each choice of the initial  product state the dynamics of a Heisenberg-coupled $N$-qubit array are governed by 
the same total Hamiltonian $H(t)=H_{XXX}+B(t)Z_{n_c}$.

Aside from the dependence of $T_{\textrm{min}}$ on the initial product state and the position of the actuator qubit, the time-dependent optimal control 
field $B_{\textrm{opt}}(t)$ found in our state-engineering scheme also depends on 
$|\psi(t=0)\rangle$ and $n_c$. This dependence is illustrated on the example of the state $|D^{4}_2\rangle$ in Fig.~\ref{fig:act-perms-pulses}. 

Another interesting feature of the obtained results for $T_{\textrm{min}}$ -- that can be gleaned from Fig.~\ref{fig:act-perms}(b) in which $N=5$ -- is that for $n_c=3$ (i.e., if the actuator is located in the center of the qubit array) our optimal-control scheme allows the dynamical generation of the Dicke state $|D^{5}_3\rangle$ only for two of the initial Hamming-weight-$3$ product states. To be more precise, the
state $|D^{5}_3\rangle$ can only be obtained starting from the states $|10101\rangle$ and $|01110\rangle$ at $t=0$. This is a special case of the previously discussed partial obstruction of subspace controllability for odd-length spin-$1/2$ chains (i.e., for odd values of $N$) with center-spin actuator [i.e., for $n_c=(N+1)/2$] (cf. Sec.~\ref{ObstructControllability}).

The state $|D^{5}_3\rangle$ is 
reflection-invariant with respect to the actuator (cf. Sec.~\ref{ObstructControllability}), i.e.,
$R|D^{5}_3\rangle = |D^{5}_3\rangle$.
Therefore, this state belongs to the six-dimensional
parity $+1$ sector $\mathcal H_3^{(+)}$ of the 
(ten-dimensional) 
three-excitation subspace $\mathcal H_3$;
its decomposition in the basis of this sector [cf. Eq.~\eqref{BasisH3pl}]
is given by
\begin{eqnarray}
|D^{5}_3\rangle
&=&\frac{1}{\sqrt{5}}
\Big( |\varphi_1^+\rangle
+ |\varphi_2^+\rangle
+ |\varphi_3^+\rangle
+ |\varphi_4^+\rangle\nonumber \Big) \\
&+&\frac{1}{\sqrt{10}}\Big(|\varphi_5^+\rangle+|\varphi_6^+\rangle\Big)\:.
\end{eqnarray}
Because the reflection-invariant product states 
$|10101\rangle\equiv|\varphi_5^+\rangle$ and $|01110\rangle\equiv|\varphi_6^+\rangle$ also belong 
to $\mathcal H_3^{(+)}$, 
the state $|D^{5}_3\rangle$ can be dynamically generated starting from one of those states at $t=0$ even for $n_c=3$,
as borne out by our numerical calculations [cf. Fig.~\ref{fig:act-perms}(b)]. By contrast to these two initial states, the remaining $8$ 
Hamming-weight-$3$ product states 
in Fig.~\ref{fig:act-perms}(b) are not in 
$\mathcal H_3^{(+)}$ and, consequently, for $n_c=3$ the state $|D^{5}_3\rangle$ cannot be obtained through dynamical evolutions of the system that start from one of those states. Needless to say, $|D^{5}_3\rangle$ can still be obtained from each of those $8$ states for any other choice of the actuator qubit (i.e., for $n_c=1,2,4$, or $5$), as also borne out by our numerical calculations.

\section{Robustness of the control scheme: Results and Discussion} \label{SchemeRobustness}
In Sec.~\ref{IdealSysResults} we discussed the optimal control-based preparation of Dicke states in idealized Heisenberg-coupled qubit arrays. It is worth noting, however, that experimental realizations of optimal-control protocols may be hampered by various real-world imperfections that ought to be taken into account when assessing the feasibility of such realizations. To this end, in the following we provide a quantitative analysis of the sensitivity of the envisioned Dicke-state preparation scheme to three different sources of imperfections, demonstrating in the process the robustness of our scheme against all three of them. To be more specific, here we quantitatively assess the effects of control-field
distortions from the optimal pulse shape  [Sec.~\ref{ControlFieldDistort}], 
control-field leakage away from the 
actuator qubit [Sec.~\ref{ControlFieldLeakage}], and control-field misalignment from its nominal direction [Sec.~\ref{ControlFieldMisalign}].

For each of the aforementioned three sources of 
real-world imperfections, the control-based part of the system dynamics is described by 
somewhat modified control Hamiltonians -- accounting for the specific source of imperfections, hence being different from $H_C(t)$ -- which we jointly denote by $H^{\textrm{imp}}_C(t)$. To quantify the effect of each of these sources of imperfections in the preparation of the Dicke state $|D^{N}_a\rangle$, we first compute numerically 
the state $|\psi_{\textrm{imp}}(t=T_{\textrm{min}})\rangle$ 
of the imperfect system at $t=T_{\textrm{min}}$ [the shortest possible state-preparation time, as obtained using dCRAB (cf. Sec.~\ref{dCRABbasedDicke})] by propagating 
the TDSE
\begin{equation}
\label{TDSE}
\frac{d}{dt}|\psi_{\textrm{imp}}(t)\rangle = -i
[H_{XXX}+ H^{\textrm{imp}}_C(t)]\:|\psi_{\textrm{imp}}(t)\rangle\:, 
\end{equation}
assuming that the initial 
state $|\psi_{\textrm{imp}}
(t = 0)\rangle$ of the system 
is the Hamming-weight-$a$ 
product state [cf. Eq.~\eqref{init_state}] previously adopted as the initial state for our control scheme and that the Hamiltonian 
$H^{\textrm{imp}}_C(t)$ derives from its ideal-system counterpart $H_C(t)$ in which the control field $B(t)$ assumes its optimal form $B_{\textrm{opt}}(t)$.

To be more specific, we propagate the TDSE in Eq.~\eqref{TDSE} up to $t=T_{\textrm{min}}$ by employing the same methodology~\cite{Tsitouras:11} as in 
the case of Eq.~\eqref{TDSEideal} (cf. Sec.~\ref{MethodsOptCtrl}).
We then obtain the Dicke-state fidelity $\mathcal{F}_{\textrm{imp}}$ of the imperfect system, i.e., the module squared of the overlap between the state $|\psi_{\textrm{imp}}
(t = T_{\textrm{min}})\rangle$ of the imperfect system at time $t=T_{\textrm{min}}$ and the target Dicke state $|D^{N}_a\rangle$:
\begin{equation}\label{impfidelity}
\mathcal{F}_{\textrm{imp}}=\big|\langle\psi_{\textrm{imp}}(t = T_{\textrm{min}})|
D^{N}_a\rangle\big|^2\:.
\end{equation} 
Finally, we quantify the effect 
of the specific source of imperfections on our 
state-preparation scheme by evaluating the 
relative deviation 
$(\mathcal{F}-\mathcal{F}_{\textrm{imp}})/\mathcal{F}
\equiv 1-\mathcal{F}_{\textrm{imp}}/\mathcal{F}$ 
of the fidelity resulting from that particular source.

\subsection{Robustness against control-field distortions 
from the optimal pulse shape} \label{ControlFieldDistort}
In quantum-technology applications, in which 
a high degree of control over the dynamics 
of quantum systems is often essential, 
it is crucial to be able to quantify 
an error resulting from random deviations 
of an external control field from its optimal values~\cite{Stojanovic:19}. Therefore, 
it is usually of pivotal importance to be able to design control pulses which -- even if not being 
optimal -- lead to an error in the relevant figure of merit (compared to the optimal value) that is smaller than some predefined threshold value, being at the same time amenable to an experimental realization.

In the Dicke-state engineering problem at hand, 
we determine the optimal control field
$B_{\textrm{opt}}(t)$ that permits the realization of a desired Dicke state in the shortest possible time (cf. Sec.~\ref{IdealSysResults}). In accordance with the above general considerations, it is pertinent to quantify the sensitivity of the state-engineering scheme under consideration against deviations from the obtained optimal control fields $B_{\textrm{opt}}(t)$. With this motivation, we consider time-dependent distortions from the optimal 
control field, which can be parametrized as~\cite{Nauth+Stojanovic:22}
\begin{equation} \label{DistortCtrlField}
\delta B(t) = \tau \sin \left(2\pi\kappa\: 
\frac{t}{T_{\textrm{min}}}\right)\:
\frac{\rm d}{{\rm d}t}B_{\textrm{opt}}(t) \:,
\end{equation}
where $\kappa$ is an integer-valued 
parameter describing the modulation 
rate of the optimal pulses; at the same time, the product of $\tau$, which has units 
of time, and the first derivative ${\rm d}B_{\textrm{opt}}(t)/{\rm d}t$ of the 
optimal control field constitutes the distortion amplitude. The total distorted control field corresponding to $\delta B(t)$ is then given by 
\begin{equation} \label{TotalDistCtrlField}
B_{\textrm{dist}}(t)\equiv B_{\textrm{opt}}(t)
+\delta B(t) \:.
\end{equation}
Despite its relatively simple form, the proposed parametrization of control-field distortions suffices to 
describe -- provided that the values 
of the parameters $\kappa$ and 
$\tau$ are chosen appropriately -- an 
almost arbitrary control-pulse shape~\cite{Nauth+Stojanovic:22}. 

\begin{figure}[t!]
\centering
\includegraphics[width=0.95\linewidth]{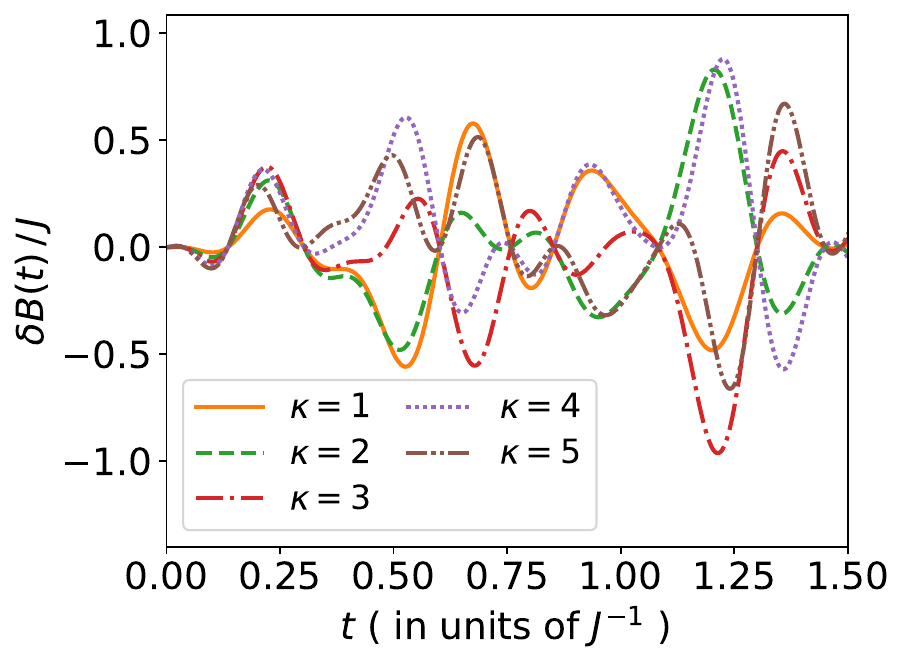}
\caption{Illustration of control-field distortions
from the optimal control-pulse shape: time-dependence of the control-field distortion $\delta B(t)$
for $\tau=0.01$ and several different values of 
the parameter $\kappa$. The results displayed 
here correspond to the state $|D^4_2\rangle$,
realized using the dCRAB formalism with $M=15$;
the optimal control-pulse shape is shown 
in Fig.~\ref{fig:TimeDepCtrl}.}
\label{fig:DistCtrlField}
\end{figure}

\begin{figure}[t!]
\centering
\includegraphics[width=0.95\linewidth]{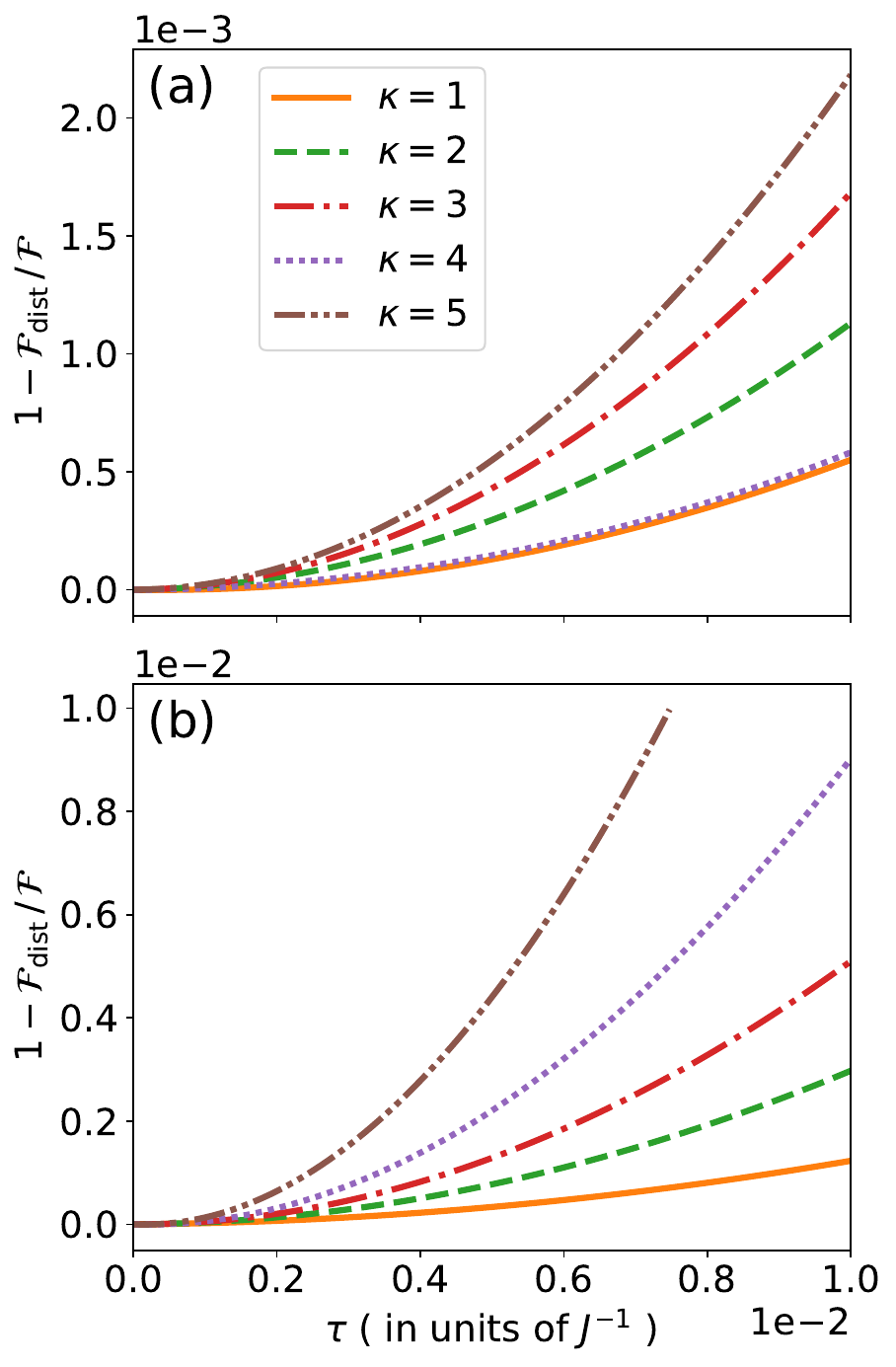}
\caption{Relative deviation from the optimal 
value of the target-state fidelity due to 
control-field distortions characterized by 
the parameters $\tau$ and $\kappa$. 
The results displayed here correspond to the states 
(a) $|W_5\rangle$, and (b) $|D^4_2\rangle$, both realized 
using the dCRAB formalism with $M=15$,
assuming that the product 
state in Eq.~\eqref{init_state}
is the initial ($t=0$) state of the system.}
\label{fig:DistFidelity}
\end{figure}

Given the character of optimal control fields $B_{\textrm{opt}}(t)$ in the state-engineering problem at 
hand -- which are varying rapidly with time (cf. Fig.~\ref{fig:TimeDepCtrl}) -- only
very small values of the parameter $\tau$ 
allow one to mimic small control-pulse distortions characteristic of present-day pulse-shaping hardware 
(cf. Sec.~\ref{IdealSysResults}). More 
specifically yet, using the above general expression for time-dependent control-field distortions $\delta B(t)$ [cf. Eq.~\eqref{DistortCtrlField}],  
it is straightforward to verify 
that $\tau\sim 0.01$ 
is the appropriate order of magnitude for the value of this parameter that allows one to emulate such small control-pulse distortions. Examples of 
control-field 
distortions $\delta B(t)$ are displayed in Fig.~\ref{fig:DistCtrlField} for 
$\tau = 0.01$ and several values of 
the parameter $\kappa$.

In the context of the Dicke-state engineering problem at hand, it is of interest to quantify to what extent the presence of small control-field distortions from the optimal
pulse shape affects our proposed scheme.
To this end, it is necessary to compute
the Dicke-state fidelity $\mathcal{F}_{\textrm{dist}}$, a special case of 
$\mathcal{F}_{\textrm{imp}}$ [cf. Eq.~\eqref{impfidelity}], corresponding to the distorted control field $B_{\textrm{dist}}(t)$; this quantity that can straightforwardly be obtained once the state $|\psi(t = T_{\textrm{min}} )\rangle$ of the system at $t = T_{\textrm{min}}$ is computed. 
This can be done in a numerically-exact fashion, namely by numerically propagating the 
TDSE~\cite{Stojanovic+Salom:19} 
of the system -- a special case of Eq.~\eqref{TDSE} in which the role of $H^{\textrm{imp}}_C(t)$ is played by
$B_{\textrm{dist}}(t)Z_1$  -- up to 
$t = T_{\textrm{min}}$. Like in the idealized (distortion-free) case, for the target state $|D^{N}_{a}\rangle$ the initial condition  
$|\psi(t=0)\rangle$ in this last dynamical equation is given by the Hamming-weight-$a$ product state 
in Eq.~\eqref{init_state}.

An example of the obtained numerical results for the relative deviation $(\mathcal{F}-\mathcal{F}_{\textrm{dist}})/\mathcal{F}\equiv 1-
\mathcal{F}_{\textrm{dist}}/\mathcal{F}$ of the 
target-state fidelity with respect to its intrinsic 
(in the absence of control-field distortions) value $\mathcal{F}$ is displayed in Fig.~\ref{fig:DistFidelity} for a range of 
values of the parameter $\tau$ and several  different choices of $\kappa$. What can be inferred from this plot is that the relative deviation in the state fidelity resulting from the control-field distortions is fairly small; this speaks in favor of the robustness of our state-engineering scheme. 

\subsection{Robustness against control-field leakage
away from the actuator qubit} \label{ControlFieldLeakage}
It is important to note that realistic control fields -- for instance, magnetic fields realized with the aid of 
micromagnets~\cite{PioroLadriere+:08} -- can never be perfectly localized. In other words, our assumption 
about the control field being confined to a single
actuator qubit constitutes an idealization. It is thus
pertinent to also briefly discuss a more realistic scenario, with a control field that also affects neighboring 
qubits as a result of control-field leakage. Describing such a scenario requires a slight generalization of our adopted control Hamiltonian $H_C(t)=B(t)Z_1$. 

Before discussing specific 
scenarios of control-field leakage, it should be pointed out that leakage effects do 
not invalidate the rationale 
behind our use of the 
local-control 
approach. Namely, as demonstrated in 
Ref.~\onlinecite{Wang++:16},
the subspace-controllability results remain valid in the presence of leakage. To be more specific, the invariant-subspace 
structure and controllability of the system remain the same as in the leakage-free case. 

Generalized control Hamiltonians, which account for the 
effects of control-field leakage, can be written 
in the form~\cite{Wang++:16} 
\begin{equation}\label{H_cLeak}
H^{cl}_C (t) =  B (t) \sum_{n=1}^{N} 
\gamma_n Z_n  \:,
\end{equation}
where -- depending on the specific type (i.e. spatial dependence) of field leakage -- 
$\gamma_n$ can assume various forms. For the sake of illustration, we consider here two different types of control-field leakage. In the case of exponential control-field leakage away from the actuator qubit $n_c$ we have
$\gamma_n = e^{-|n-n_c|/\zeta}$, where $\zeta$ is the parameter that measures the extent of exponential leakage; note that $\zeta =0$ in the idealized (leakage-free) case. Likewise, in the case of Gaussian 
control-field leakage $\gamma_n=e^{-(n-n_c)^2/(2\sigma^2)}$, where $\sigma$ is the parameter quantifying this type of leakage; once again, $\sigma=0$ indicates the complete absence of leakage.
In the special $n_c=1$ case (i.e., when the first
qubit in the array plays the role of the actuator 
qubit), which is of particular interest here, the above expressions for $\gamma_n$ go over into $\gamma_n = e^{-(n-1)/\zeta}$ and $\gamma_n=e^{-(n-1)^2/(2\sigma^2)}$, respectively.

In the following, we quantify the effects of 
control-field leakage in the state-engineering 
problem at hand by evaluating Dicke-state fidelities 
in the presence of leakage. In other words, we determine
those fidelities with $B(t)=B_{\textrm{opt}}(t)$ (i.e., for the optimal control fields computed in the leakage-free case [cf. Sec.~\ref{IdealSysResults}]), assuming at the same time that the evolution of the system is
governed by the generalized control Hamiltonian of Eq.~\eqref{H_cLeak}, with $\gamma_n$ corresponding to either exponential- or Gaussian leakage. To this end, for each desired Dicke (or, in the special case, $W$) state we propagate the corresponding TDSE~\cite{Stojanovic+Salom:19} [with the initial state $|\psi(t=0)\rangle$ given by Eq.~\eqref{init_state} in the case of the target
state $|D^{N}_{a}\rangle$] -- a special case of Eq.~\eqref{TDSE}, with the role of $H^{\textrm{imp}}_C(t)$ being played by
$H^{cl}_C(t)$ [cf. Eq.~\eqref{H_cLeak}] -- 
up to $t=T_{\textrm{min}}$, i.e., 
the time corresponding to the time-optimal, dCRAB-based realization of the desired Dicke (or $W$) state (cf. Sec.~\ref{IdealSysResults}). For a range of values of the parameters $\zeta$ and $\sigma$ (for exponential- and Gaussian leakage, respectively), we then compute the fidelity $\mathcal{F}_{\textrm{leak}}$ -- a special case of $\mathcal{F}_{\textrm{imp}}$ [cf. Eq.~\eqref{impfidelity}] -- as the module squared of the overlap between the target
Dicke (or $W$) state and the actual state 
$|\psi_{\textrm{leak}} (t=T_{\textrm{min}})\rangle$ of the system at $t=T_{\textrm{min}}$ in the presence of leakage, obtained as described above. 

\begin{figure}[t!]
\centering
\includegraphics[width=0.95\linewidth]{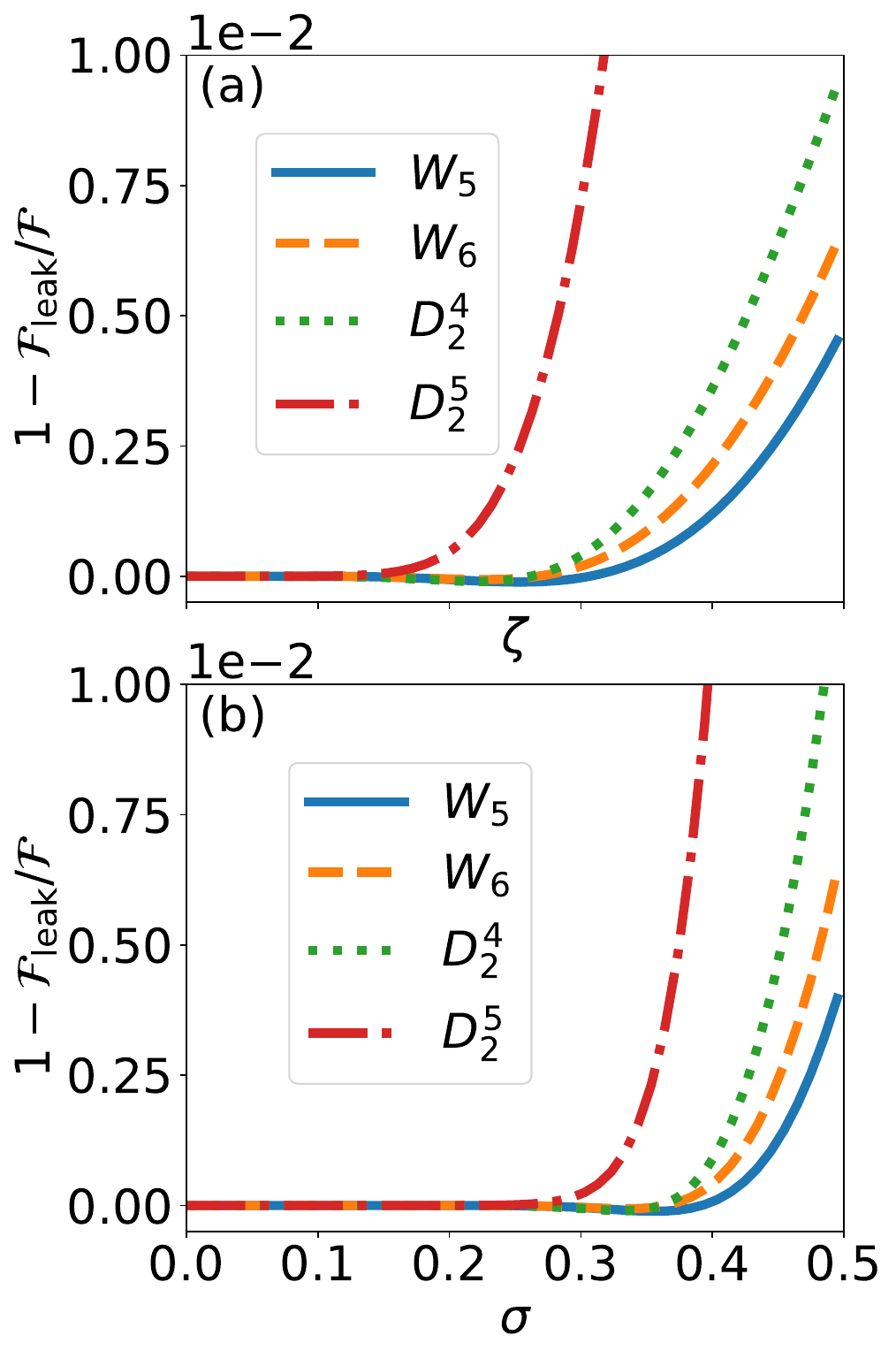}
\caption{Relative deviations $1-
\mathcal{F}_{\textrm{leak}}/\mathcal{F}$ of the Dicke- and $W$ state fidelities $\mathcal{F}_{\textrm{leak}}$ in the presence of control-field leakage away from the actuator qubit. The intrinsic optimal values $\mathcal{F}$ of the state fidelity are obtained using the dCRAB formalism with $M=15$, assuming that the product state of Eq.~\eqref{init_state}
is the initial ($t=0$) state of the system. The results shown here correspond to the cases of (a) exponential leakage, described by the parameter $\zeta$, and (b) Gaussian leakage, described by the parameter $\sigma$.}
\label{fig:error-leak}
\end{figure}

We choose the range $[0,0.5]$ of values 
for the parameters $\zeta$ and $\sigma$. In particular, for the largest considered value $\zeta=\sigma = 0.5$ of those parameters, the control-field leakage on the nearest neighbor ($n=2$) of the actuator qubit amounts -- for both exponential- and 
Gaussian control-field 
leakage -- to $e^{-2}\approx 13.5\:\%$ of the control-field 
magnitude on the actuator qubit. On the other hand, for their median value $\zeta=\sigma = 0.25$, one has $e^{-4}\approx 1.8\:\%$ of the original magnitude in the exponential- and $e^{-8}\approx 0.03\:\%$ in the Gaussian-leakage case. 

The effect of control-field leakage 
on the $W$- and Dicke-state 
fidelities -- as quantified by
the relative deviation $(\mathcal{F}-\mathcal{F}_{\textrm{leak}})/\mathcal{F}\equiv 1-
\mathcal{F}_{\textrm{leak}}/\mathcal{F}$ of the 
target-state fidelity with respect to its intrinsic 
(in the absence of leakage) value $\mathcal{F}$ -- is illustrated in Fig.~\ref{fig:error-leak} for both exponential- and Gaussian leakage. 
What can be inferred from the obtained results is that the dependence of $1-
\mathcal{F}_{\textrm{leak}}/\mathcal{F}$ on $\zeta$ (or $\sigma$) is very similar for all target states considered. In addition, an important quantitative conclusion can also be drawn from Fig.~\ref{fig:error-leak}; namely, for a reasonable range 
of values of parameters $\zeta$ and $\sigma$, the resulting fidelities in the presence of leakage ($\mathcal{F}_{\textrm{leak}}$) are reduced 
from the corresponding intrinsic values 
($\mathcal{F}$) by an amount of the order of 
$10^{-3}$, i.e., $0.1\%$. This 
finding convincingly demonstrates the robustness of our optimal local-control scheme for Dicke-state generation against 
control-field leakage away from the actuator qubit.

\subsection{Robustness against control-field misalignment from its nominal direction} \label{ControlFieldMisalign}
Because our scheme for generating Dicke states is predicated on a fixed spatial direction of the external control field -- namely, that of the $z$ axis -- it is important to demonstrate the robustness 
of this scheme against 
small control-field misalignments from this nominal direction. In other words, it is pertinent to assume that the direction of the control field is not ideally aligned with the $z$ axis, but is instead given by the unit vector 
\begin{equation}\label{UnitVecCtrl}
\hat{\mathbf{n}}_{\textrm B}\equiv
(\sin\theta_{\textrm B}\cos\phi_{\textrm B},\sin\theta_{\textrm B}\sin\phi_{\textrm B},
\cos\theta_{\textrm B})^{\textrm{T}} \:,
\end{equation}
where $\theta_{\textrm B}$ and 
$\phi_{\textrm B}$ are the spherical 
polar- and azimuthal angles. Under this last assumption, the control part of the total system Hamiltonian 
assumes the form $H^{cm}_C (t)=B(t)\hat{\mathbf{n}}_{\textrm B}\cdot \mathbf{X}_1$ [cf. Eq.~\eqref{defineXYZ_n}], instead of being given by 
$H_C (t) =  B (t) Z_1$. Using 
Eq.~\eqref{UnitVecCtrl}, the above succinct form of $H^{cm}_C (t)$ can be recast as
\begin{eqnarray}\label{H_cMis}
H^{cm}_C (t) &=& B(t)\times(\sin\theta_{\textrm B}\cos\phi_{\textrm B} X_1 
+ \sin\theta_{\textrm B}\sin\phi_{\textrm B} Y_1  
\nonumber\\
&+&\cos\theta_{\textrm B} Z_1) \:.
\end{eqnarray}
In particular, a small misalignment of the control field from its nominal ($z$) direction corresponds to 
the small values $\theta_{\textrm B}\ll \pi$ of the spherical polar angle.

\begin{figure}[t!]
\centering
\includegraphics[width=0.95\linewidth]{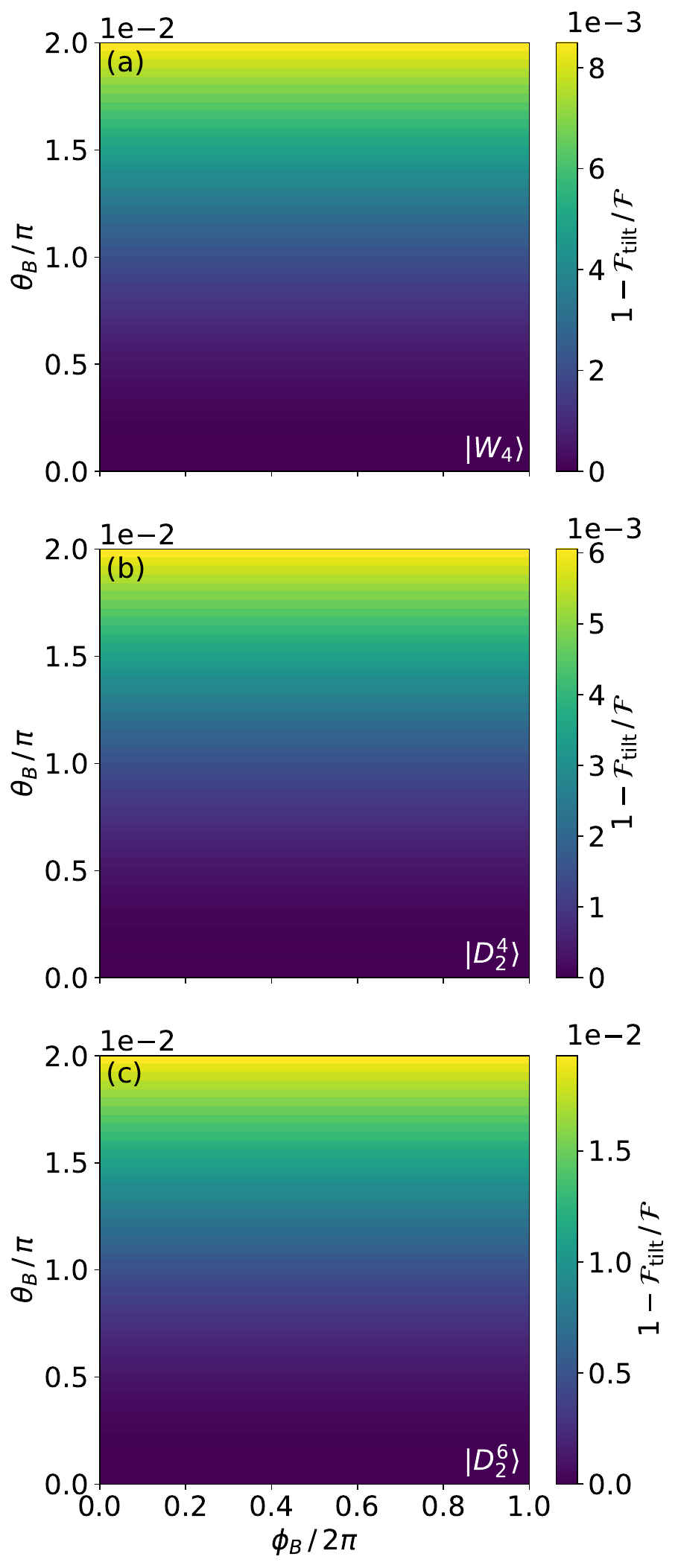}
\caption{Relative deviations $1-
\mathcal{F}_{\textrm{tilt}}/\mathcal{F}$ of the Dicke- and $W$ state fidelities
$\mathcal{F}_{\textrm{tilt}}$ resulting from control-field misalignment, characterized by the spherical polar ($\theta_{\textrm B}$) and azimuthal
($\phi_{\textrm B}$) angles, from their intrinsic optimal values $\mathcal{F}$; the latter are obtained using the dCRAB formalism with $M=15$,
assuming that the product 
state of Eq.~\eqref{init_state}
is the initial ($t=0$) state of the system. The results shown here correspond to the states (a) $\lvert W_4\rangle$, (b) $\lvert D^4_2\rangle$, and (c) $\lvert D^6_2\rangle$.}
\label{fig:error-tilt}
\end{figure}

In what follows, we 
quantify the effects 
of control-field misalignment in the 
state-engineering problem under consideration
by evaluating Dicke-state fidelities 
assuming that the control Hamiltonian 
of the system is given by the one in 
Eq.~\eqref{H_cMis}. To that end, for 
each of the desired Dicke (or, in the special case, $W$) states we propagate the corresponding 
TDSE [with the initial state $|\psi(t=0)\rangle$ given by Eq.~\eqref{init_state} in the case of the target
state $|D^{N}_{a}\rangle$] up to $t=T_{\textrm{min}}$, i.e., 
the time corresponding to the time-optimal, dCRAB-based realization of the desired Dicke (or $W$) state (cf. Sec.~\ref{IdealSysResults}); this TDSE is a special case of Eq.~\eqref{TDSE} in which the role of $H^{\textrm{imp}}_C(t)$ is played by
$H^{\textrm{cm}}_C(t)$. For a range of small values of $\theta_{\textrm B}$ and a range of values of $\phi_{\textrm B}$, 
we then evaluate the fidelity $\mathcal{F}_{\textrm{tilt}}$ -- a special case of 
$\mathcal{F}_{\textrm{imp}}$ in  Eq.~\eqref{impfidelity} -- which is given by the module squared of the overlap between the target
Dicke (or $W$) state and the actual state 
$|\psi_{\textrm{cm}} (t=T_{\textrm{min}})\rangle$ of the system at $t=T_{\textrm{min}}$ in the presence of control-field misalignment, obtained as described above. By analogy to the other sources of imperfections (cf. Secs.~\ref{ControlFieldDistort} and \ref{ControlFieldLeakage}), we quantify 
the effect of control-field misalignment by computing the relative change $(\mathcal{F}-\mathcal{F}_{\textrm{tilt}})/\mathcal{F}\equiv 1-
\mathcal{F}_{\textrm{tilt}}/\mathcal{F}$ of the target-state fidelity with respect to its intrinsic (in the absence of control-field misalignment) value $\mathcal{F}$.

The numerical results obtained for three different target Dicke states are shown in Fig.~\ref{fig:error-tilt} for small values of the polar angle $\theta_{\textrm{B}}$ -- with the maximal variation chosen to be as large as $2\%$ 
of its maximal value $\pi$ -- and the whole range $[0,2\pi)$ of values  for the azimuthal angle
$\theta_{\textrm{B}}$. What can be inferred 
from Fig.~\ref{fig:error-tilt} is that the corresponding relative change in the state fidelity 
is below $1\%$ for the states 
$|W_4\rangle$ and $|D^{4}_2\rangle$, while in the case of the state $|D^{6}_2\rangle$
it is around $1.8\%$. These results convincingly 
demonstrate the robustness of our envisioned state-preparation scheme against small control-field misalignment.

\section{Summary and Conclusions} \label{SummConcl}
To summarize, the feasibility of time-efficient analog (single-shot) realizations of Dicke states was investigated in this paper for qubit arrays with the isotropic, nearest-neighbor, always-on Heisenberg coupling between qubits 
and a single Zeeman-type local (i.e., acting on a single actuator qubit) control in the $z$ direction. The theoretical underpinning for the state-control scheme that permits the generation of Dicke states $|D^{N}_{a}\rangle$ 
($a=1,\ldots,N-1$) starting from a generic Hamming-weight-$a$ product state is provided by an already proven, general result in the realm of Lie-algebraic controllability of interacting spin-$1/2$ chains (qubit arrays). This result guarantees 
the subspace controllability of 
Heisenberg-coupled qubit arrays on any subspace of the total $N$-qubit Hilbert space that is characterized by a fixed number of excitations (i.e., a fixed Hamming weight). More specifically yet, this result implies -- if a $Z$ control is applied to a single qubit -- that a time-dependent control field can be found that allows one to realize the Dicke state with the desired excitation number $a$ starting from an arbitrary product (separable) state with the same number of excitations. 

The problem of finding the appropriate time-dependence of the control field acting on an actuator qubit for engineering Dicke states -- including 
$W$ states as their special, single-excitation 
case -- in the shortest possible time was addressed here for linear arrays with up to $9$ qubits using methods of quantum optimal control. More specifically yet, the search for optimal, smoothly-varying control fields that enable the realization of both $W$- and genuine Dicke states at the quantum speed limit was carried out using the dCRAB algorithm, which we implemented in combination with advanced methods of global optimization. In particular, based on our obtained numerical results, 
we showed that the shortest possible times required for high-fidelity realizations of $W$- and two-excitation Dicke states in a Heisenberg-coupled 
$N$-qubit array in this single-control setting scale 
as $\mathcal{O}(N^{2.08})$ and $\mathcal{O}(N^{1.78})$, respectively. To demonstrate the practical viability of our proposed state-engineering scheme, its sensitivity to various real-world imperfections -- such as control-field distortions from the optimal pulse shape, control-field leakage away from the actuator qubit, and 
control-field misalignment from its nominal $z$ direction  -- was also investigated, yielding favorable results. 

To conclude, the present work underscores the potential usefulness of optimal-control schemes for time-efficient and robust engineering of
highly-entangled multiqubit states of interest 
for quantum-technology applications. In addition, it points to the utility of the local-control approach -- i.e.,
the use of minimal control resources that permit 
the realization of a desired 
quantum-control task -- in this 
context. Experimental realizations of such approaches, especially those based on solid-state qubits with always-on interactions, are clearly 
called for. 

\begin{acknowledgments}
This research has received funding (V.M.S.) from the European Union's Horizon Europe programme 
HORIZON-CL4-2021-DIGITAL-EMERGING-01-30 via the project 101070144 (EuRyQa) and from the French National Research Agency under the Investments of the Future Program projects ANR-21-ESRE-0032 (aQCess).
We acknowledge funding from the Horizon Europe programme HORIZON-CL4-2022-QUANTUM-02-SGA via the project 101113690 PASQuanS 2.1, as well as by Germany’s Excellence Strategy – Cluster of Excellence Matter and Light for Quantum Computing (ML4Q2) EXC 2004/2 – 390534769 (T. C. and A. M).
\end{acknowledgments}

\end{document}